\documentclass[twocolumn,twocolappendix]{aastex63}
\usepackage{amsmath}
\usepackage{xcolor}
\usepackage[normalem]{ulem}
\definecolor{dgreen}{RGB}{26,148,49}
\definecolor{xlinkcolor}{cmyk}{1,1,0,0}
\hypersetup{linkcolor=xlinkcolor,citecolor=xlinkcolor,urlcolor=xlinkcolor}

\usepackage{lineno}

\DeclareRobustCommand{\ion}[2]{%
\relax\ifmmode
\ifx\testbx\f@series
{\mathbf{#1\,\mathsc{#2}}}\else
{\mathrm{#1\,\mathsc{#2}}}\fi
\else\textup{#1\,{\mdseries\textsc{#2}}}%
\fi}

\shorttitle{CNIB Tomography with SPHEREx}
\shortauthors{Cheng \& Chang}

\begin{document}

\title{Cosmic Near-infrared Background Tomography with SPHEREx Using Galaxy Cross-correlations}

\author[0000-0002-5437-0504]{Yun-Ting Cheng}
\address{California Institute of Technology, 1200 E. California Boulevard, Pasadena, CA 91125, USA}
\email{ycheng3@caltech.edu}

\author[0000-0001-5929-4187]{Tzu-Ching Chang}
\address{California Institute of Technology, 1200 E. California Boulevard, Pasadena, CA 91125, USA}
\address{Jet Propulsion Laboratory, California Institute of Technology, 4800 Oak Grove Drive, Pasadena, CA 91109, USA}

\begin{abstract}
The extragalactic background light (EBL) consists of integrated light from all sources of emission throughout the history of the universe. At near-infrared wavelengths, the EBL is dominated by stellar emission across cosmic time; however, the spectral and redshift information of the emitting sources is entangled and cannot be directly measured by absolute photometry or fluctuation measurements. Cross-correlating near-infrared maps with tracers of known redshift enables EBL redshift tomography, as EBL emission will only correlate with external tracers from the same redshift. Here, we forecast the sensitivity of probing the EBL spectral energy distribution as a function of redshift by cross-correlating the upcoming near-infrared spectro-imaging survey, SPHEREx, with several current and future galaxy redshift surveys. Using a model galaxy luminosity function, we estimate the cross power spectrum clustering amplitude on large scales, and forecast that the near-infrared EBL spectrum can be detected tomographically out to $z\sim 6$. We also predict a high-significance measurement ($\sim 10^2$--$10^4\sigma$) of the small-scale cross-power spectrum out to $z\sim 10$. The amplitudes of the large-scale cross power spectra can constrain the cosmic evolution of the stellar synthesis process through both continuum and the line emission, while on the nonlinear and Poisson noise scales, the high-sensitivity measurements can probe the mean spectra associated with the tracer population across redshift. \footnote{\textcircled{c} 2021. All rights reserved.}
\end{abstract}

\section{Introduction}
The extragalactic background light (EBL) is the aggregate light from all sources of emission across cosmic time. EBL measurements have been made from gamma-rays to radio \citep[see, e.g.,][ for recent reviews]{2016RSOS....350555C,2019ConPh..60...23M}. At near-infrared wavelengths, the EBL is mostly produced by redshifted ultraviolet and optical stellar emission, and thus carries essential information on the history of stellar synthesis processes in our universe. However, observations also suggest other sources of the near-infrared EBL, including diffuse light in the dark matter halos from stripped stars, sometimes referred to as intrahalo light \citep[IHL; ][]{2012Natur.490..514C,2014Sci...346..732Z,2021arXiv210303882C}, and early formation of stars and galaxies from the epoch of reionization and cosmic dawn \citep{2005Natur.438...45K,2011ApJ...742..124M,2012ApJ...753...63K,2015NatCo...6.7945M}. Possible near-infrared EBL emission from Population III stars and accretion disks around their stellar mass black holes from $z>7$ \citep{2018ApJS..234...41W}, direct collapse black holes from the dark ages \citep{2013MNRAS.433.1556Y,2014MNRAS.440.1263Y} and the decay of axion-like particles as dark matter candidates \citep[e.g., ][]{2016ApJ...825..104G,2020arXiv201209179C} have also been proposed.

Extensive studies have attempted to probe the near-infrared EBL through different methods. Galaxy counts measures emission from resolved galaxies and the observations are extrapolated to estimate contributions from faint galaxies below the detection limit \citep{2010ApJ...723...40K,2011MNRAS.410.2556D,2012ApJ...752..113H,2016ApJ...827..108D,2020arXiv201203035S,2021arXiv210212323K}.  This sets a lower bound to the integrated galaxy light (IGL) component of EBL, as EBL from diffuse emission or sources not associated with galaxies is not included. Direct measurements using absolute photometry capture all emission in the EBL \citep{2007ApJ...666..663B, 2007ApJ...666...34L,2013PASJ...65..121T,2015ApJ...807...57M,2015ApJ...811...77S,2017ApJ...839....7M,2017NatCo...815003Z,2020arXiv201103052L,2020ApJ...901..112S}, however, absolute photometry is challenging since systematic errors and foreground contamination have to be tightly controlled. Fluctuation analysis is an alternative approach that is less susceptible to foreground contamination (e.g. zodiacal light, Galactic cirrus), since foregrounds and the EBL signal have distinct spatial and spectral correlation features  \citep{2005Natur.438...45K,2007ApJ...666..658T,2011ApJ...742..124M,2012Natur.490..514C,2012ApJ...753...63K,2014Sci...346..732Z,2015NatCo...6.7945M,2015ApJ...807..140S,2019PASJ...71...82K,2019PASJ...71...88M}; nevertheless inferring the absolute intensity of the EBL from fluctuation measurements depends on model assumptions. Another limitation of absolute photometry and fluctuation measurements is the inferred EBL redshift resolution. As the measured intensity is a projection of all emission along the line of sight, one way to infer the redshift dependency of the EBL is to measure the opacity of gamma-ray photons from individual blazars. Near-infrared photons along the line of sight will interact with gamma-ray photons by pair production, and thus the redshift composition of EBL can be constrained by the absorption features in blazar spectra \citep[e.g., ][]{2006Natur.440.1018A,2007A&A...475L...9A,2008Sci...320.1752M,2010ApJ...723.1082A,2012Sci...338.1190A,2017A&A...606A..59H,2018ApJS..237...32A,2019ApJ...885..150A,2019MNRAS.486.4233A,2020MNRAS.494.5590A}. This method enables a redshift tomography of the EBL by observing blazars from different distances. However, these estimates have low spectral resolution and depend on the assumption of intrinsic blazar spectra.

As the EBL intensity contains a mixture of emitting sources from all redshifts, and different sources have distinct spectral features, it will be insightful to decompose the EBL signal by its redshift and spectral dependencies. Cross-correlation provides a path to perform EBL tomography while retaining spectral information. By cross-correlating with external sources (e.g., a galaxy catalog) that trace the underlying large-scale structure at a certain redshift, one can extract the IGL and other components associated with the large-scale structure, such as the IHL, on large scales. The cross-correlation technique has previously been proposed and applied to derive the redshift distribution of photometric catalogs \citep{2008ApJ...684...88N,2013MNRAS.433.2857M,2013arXiv1303.4722M}, 
perform redshift tomography of broadband imaging surveys \citep{2015MNRAS.446.2696S,2019ApJ...870..120C,2019ApJ...877..150C}, and probe the cosmic infrared background \citep{2014A&A...570A..98S} and the thermal or kinematic Sunyaev--Zel'dovich signal from the cosmic microwave background \citep[e.g., ][]{2016PhRvL.117e1301H, 2020ApJ...902...56C}. Cross-correlation has also been proposed to probe the direct collapse black holes \citep{2013ApJ...769...68C,2016ApJ...832..104M} and the axion decay \citep{2018PhRvD..98f3524C} in the optical and near-infrared EBL. In line intensity mapping \citep[see ][for a review]{2017arXiv170909066K}, where the emission from a certain spectral line is used to trace the three-dimensional large-scale structure, cross-correlation is also a powerful technique to measure line emission from 21 cm \citep[e.g., ][]{2010Natur.466..463C,2013ApJ...763L..20M}, CO \citep[e.g., ][]{2013ApJ...768...15P}, [\ion{C}{ii}] \citep[e.g., ][]{2018MNRAS.478.1911P,2019MNRAS.489L..53Y}, and Ly$\alpha$ \citep[e.g., ][]{2016MNRAS.457.3541C,2018MNRAS.481.1320C} with a much reduced impact from foreground and systematics contaminations.

SPHEREx \citep{2014arXiv1412.4872D, 2018arXiv180505489D} is an approved NASA MIDEX mission that will carry out an all-sky spectro-imaging survey at near-infrared wavelengths (see Sec.~\ref{S:SPHEREx} for details). SPHEREx will achieve unparalleled spectral resolution, sensitivity, and sky coverage for spectrally mapping the near-infrared sky, which provides an exceptional dataset for studying EBL through fluctuation analysis.

Next generation spectroscopic and photometric galaxy surveys will come online in the next few years, including DESI \citep{2016arXiv161100036D}, the Rubin Observatory LSST \citep{2009arXiv0912.0201L}, Euclid \citep{2018LRR....21....2A}, and the Roman Space Telescope \citep{2015arXiv150303757S}. These surveys will provide galaxy catalogs to an unprecedented depth and synergy with the SPHEREx spectro-imaging dataset. In this work, we forecast constraints on EBL enabled by cross-correlations of SPHEREx and upcoming galaxy surveys. SPHEREx spectro-images encode the spatial and spectral information of the EBL, whereas galaxies trace the three-dimensional cosmic structures with redshift information. With cross-correlation, we perform redshift tomography of the EBL spectrum, which constrains the redshift composition and spectral response of the integrated background emission, as well as its spatial scale dependence. These measurements primarily consist of the IGL, and will provide valuable insight on the cosmic star formation and stellar mass history. Furthermore, information on the IHL or other possible EBL emitting sources that trace the large-scale structure (e.g. the decaying dark matter candidates) can also be revealed by the cross-correlation measurements.

A recent study by \citet{2021arXiv210400017S} investigated the ultraviolet to optical EBL constraints from cross-correlation, using images from the future ultraviolet survey Cosmological Advanced Survey Telescope for Optical-ultraviolet Research \citep[CASTOR; ][]{2019clrp.2020...18C} for lower redshift, and SPHEREx for the epoch of reionization. Their study focuses on the UV broadband emission and the Ly$\alpha$ line. In this work, we conduct a comprehensive forecast of constraining the EBL spectra from SPHEREx with all accessible wavelength and redshift ranges.

This paper is organized as follows. We first introduce the redshift tomography formalism of cross power spectrum estimation in Sec.~\ref{S:Formalism}. We model components of the EBL signal in Sec.~\ref{S:models}. Sec.~\ref{S:xcorr_forecast} details the evaluation metrics and the fiducial cases we choose to present the results. Sec.~\ref{S:SPHEREx} introduces SPHEREx, and Sec.~\ref{S:data} describes the galaxy surveys considered in this work. The results are presented in Sec.~\ref{S:results}. Discussion and further science interpretations are given in Sec.~\ref{S:Discussion} and Sec.~\ref{S:Interpretation}, respectively. Sec.~\ref{S:Conclusion} concludes the paper. Throughout this work, we assume a flat $\Lambda$CDM cosmology with $n_s=0.97$, $\sigma_8=0.82$, $\Omega_m=0.26$, $\Omega_b=0.049$, $\Omega_\Lambda=0.69$, and $h=0.68$, consistent with the measurement from Planck \citep{2016A&A...594A..13P}. All fluxes are quoted in the AB magnitude system.

\section{Formalism}\label{S:Formalism}
We detail the analytical expression for a cross power spectrum, and its error estimation, between a continuous density field (such as a diffuse intensity map) and a discrete density tracer sample (such as a galaxy catalog). An intensity field $I$ and a galaxy tracer field $g$ can each be expanded in terms of spherical harmonics $Y_{\ell m}$ as
\begin{equation}
\delta I\left(\hat{n}\right)=\sum_{\ell,m} a^I_{\ell m}Y_{\ell m}\left(\hat{n}\right),
\end{equation} 
and
\begin{equation}
\delta g\left(\hat{n}\right)=\sum_{\ell,m} a^g_{\ell m}Y_{\ell m}\left(\hat{n}\right).
\end{equation}

The angular cross power spectrum of these two fields,  $C_{\ell}^{\rm x}$, binned in a width of $\Delta\ell$, is given by
\begin{equation}
C_\ell^{\rm x}=\left \langle \frac{1}{\left (2\ell+1  \right )f_{\rm sky}}\sum_{m=-\ell}^{\ell}\left (a^I_{\ell m}  \right )^*a^g_{\ell m}\right \rangle_{\ell\in\Delta\ell},
\end{equation}
with a variance of
\begin{equation}\label{E:delta_Clx}
\left ( \delta C_\ell^{\rm x} \right )^2 = \frac{1}{N_\ell}\left [(C_{\ell}^{\rm x})^2+C_{\ell}^I C_{\ell}^g \right ],
\end{equation}
where $N_\ell = \Delta \ell\left ( 2\ell +1 \right )f_{\rm sky}$ is the number of $\ell$ modes in the band power spectrum at $\ell$ with a width of $\Delta\ell$, and $f_{\rm sky}$ is the fraction of sky area used in the cross-correlation. $C_{\ell}^g$, $C_{\ell}^I$, and $C_{\ell}^{\rm x}$ are the galaxy auto power spectrum, intensity field auto power spectrum, and their cross power spectrum, respectively. 

Eq.~\ref{E:delta_Clx} is the Gaussian variance that assumes all multipole modes are independent. This is a good approximation on large scales, but on small scales in the Poisson regime, a non-Gaussian contribution to the covariance needs to be accounted for. Therefore, we also consider the trispectrum contribution to the variance. The cross power spectrum covariance is thus
\begin{equation}\label{E:Cov}
{\rm Cov}\left ( C_\ell^{\rm x},C_{\ell'}^{\rm x} \right ) = \delta^K_{\ell\ell'} \left ( \delta C_\ell^{\rm x} \right )^2 + \frac{\mathcal{T}^{ gIgI}(\ell, \ell')}{\Omega_{\rm sur}},
\end{equation}
where $\delta^K$ is the Kronecker delta, $\Omega_{\rm sur}=4\pi f_{\rm sky}$ is the survey area in solid angle, and $\mathcal{T}^{gIgI}(\ell,\ell')$ is the trispectrum of the galaxy and intensity field \citep{2018PhRvD..97l3539S}.

\def\Clg{$C_\ell^g$}
\subsection{Galaxy Sample Auto Power Spectrum \Clg}\label{S:Clg}
We describe a galaxy auto power spectrum as comprised of clustering and (shot) Poisson noise terms:
\begin{equation}\label{E:Clg}
C_\ell^g = C_{\ell, \rm clus}^g+C_{\ell, \rm P}^g.
\end{equation}
Using the Limber approximation, the clustering term can be expressed as
\begin{equation}
C_{\ell, \rm clus}^g=\int dz \frac{H(z)}{c} \frac{f^g(z)f^g(z)}{\chi^2(z)}b_g^2(z)P_{\rm m}\left(k=\frac{\ell+\frac{1}{2}}{\chi(z)},z\right),
\end{equation}
where $H$, $\chi$, $c$, and $f^g(z)$ are the Hubble parameter, co-moving distance, speed of light, and the galaxy selection function, respectively. $P_m$ is the matter power spectrum. In this work, we focus on the linear clustering regime and restrict our analysis to large scales at $k<0.2$ $h$ Mpc$^{-1}$. We assume $b_g$ is a scale-independent linear bias. In practice, with an assumed cosmological model, $b_g$ can be measured from the amplitude of the galaxy auto spectrum. We consider the limit where galaxies are from a narrow redshift bin ($\Delta z^g\ll 1$) centered at $z^g$, and use the following approximation:
\begin{equation}
f^g(z) =
\begin{cases}
\frac{1}{\Delta z^g}  \quad \quad \text{if}  \quad z_g-\frac{\Delta z^g}{2} < z < z_g+\frac{\Delta z^g}{2}\\
0 \quad \quad  \quad \text{Otherwise},
\end{cases}
\end{equation}
and
\begin{equation}
C_{\ell, \rm clus}^g=\frac{1}{\Delta z^g} \frac{H(z^g)}{c\chi^2(z)}b_g^2(z)P_{\rm m}(k=\frac{\ell+\frac{1}{2}}{\chi(z^g)},z^g).
\end{equation}

The galaxy Poisson noise power is the reciprocal of the surface number density $dN_g/d\Omega$,
\begin{equation}\label{E:Clg_shot}
C_{\ell, \rm P}^g=\left ( \frac{dN_g}{d\Omega} \right )^{-1} = \left ( \Delta z^g\frac{dN_g}{dzd\Omega} \right )^{-1},
\end{equation}
where the second equality uses the same assumption of a narrow redshift bin.

\def\ClI{$C_\ell^I$}
\subsection{Intensity Field Auto Power Spectrum \ClI}\label{S:ClI}

The angular auto power spectrum of an intensity field includes contributions from clustering, Poisson noise, and instrument noise,
\begin{equation}\label{E:ClI}
C_\ell^I = C_{\ell, \rm clus}^I+C_{\ell, \rm P}^I + C_{\ell,n}^I.
\end{equation}
Using the Limber approximation, the clustering power spectrum of the intensity map at frequency $\nu$ is
\begin{equation}
\begin{split}
C_{\ell, \rm clus}^I&=\int dz \frac{H(z)}{c} \frac{f^I(z)f^I(z)}{\chi^2(z)}
b_I^2(z,\nu)\\
&\cdot \left[\frac{d(\nu I_\nu)(z,R(\nu))}{dz} \right]^2P_{\rm m} \left(k=\frac{\ell+\frac{1}{2}}{\chi(z)},z \right),
\end{split}
\end{equation}
where $b_I$ is the large-scale bias, and $f^I$ the redshift selection function. We take $f^I(z)=1$ for simplicity as the intensity field is composed of emission over a large redshift range. The term $\nu I_\nu(z,R(\nu))$ is the intensity of emitting sources at redshift $z$ observed by a filter with frequency response function $R(\nu)$, i.e.,
\begin{equation}
\frac{d(\nu I_\nu)}{dz}(z,R(\nu)) = \frac{1}{\int d\nu R(\nu)}\int d\nu R(\nu)\frac{d(\nu I_\nu)}{dz}(z,\nu),
\end{equation}
where $\nu I_\nu(z,\nu)$; (in units of nW m$^{-2}$ sr$^{-1}$) is the intrinsic spectral energy distribution of the EBL from sources at redshift $z$ observed at frequency $\nu$. In this work, we consider filters with a narrow spectral width ($\Delta\nu/\nu\ll 1$), and use the approximation:
\begin{equation}
\frac{d(\nu I_\nu)}{dz}(z,R(\nu)) = \frac{d(\nu I_\nu)}{dz}(z,\nu).
\end{equation}
The EBL is the aggregate intensity from all sources across cosmic time, defined as 
\begin{equation}
\nu I_\nu(\nu) = \int dz \frac{d(\nu I_\nu)}{dz}(z,\nu).
\end{equation}
$d(\nu I_\nu)/dz$ is the redshift-dependent EBL intensity in the same units as $\nu I_\nu$. In the optical to near-infrared wavelengths, the emitting sources contributing to the measured intensity include galaxies, quasars, and stars, from all luminosity ranges and from all redshifts \citep[e.g.,][]{2016RSOS....350555C,2019ConPh..60...23M}.

We model the integrated galaxy light (IGL) as the main emitting source of the EBL, which can be considered as a lower bound to $\nu I_\nu(z,\nu)$. We model the continuum and spectral line emission of the IGL spectrum separately,
\begin{equation}
\begin{split}
\left.\frac{d(\nu I_\nu)}{dz}\right|_{\rm IGL}(z,\nu)&=\left.\frac{d(\nu I_\nu)}{dz}\right|_{\rm cont}(z,\nu)\\
&+\sum_{\rm line}\left.\frac{d(\nu I_\nu)}{dz}\right|_{\rm line}(z,\nu).
\end{split}
\end{equation}

The continuum can be expressed in terms of a volume-averaged galaxy emissivity, consisting of the galaxy luminosity function $\Phi(m,z,\nu_{\rm rf})=dN/dV/dm$, where $N$ is the galaxy number count, $V$ is the co-moving volume, $m$ is the AB magnitude, and $\nu_{\rm rf}$ is the rest-frame frequency. The spectrum of the continuum intensity can be written as
\begin{equation}\label{E:dnuInudz_conti}
\begin{split}
\left.\frac{d(\nu I_\nu)}{dz}\right|_{\rm cont}(z,\nu)&=\int_{m_{\rm th}(z,\nu)}^{\infty}dm \Phi(m,z,\nu(1+z))\\
&\cdot\nu F_\nu(m)\frac{d\chi}{dz}(z)D_A^2(z),
\end{split}
\end{equation}
where $F_{\nu}(m)=3631\times 10^{m/2.5}$ Jy is the specific flux density at magnitude $m$, $d\chi/dz=c/H(z)$, and $D_A$ is the co-moving angular diameter distance, which equals the co-moving distance in a flat universe. The lower bound of the magnitude integration, $m_{\rm th}(z,\nu)$, is the masking magnitude threshold, below which sources are masked in the diffuse intensity map. Masking reduces the foreground contributions to the cross power spectrum Poisson noise, as discussed in Sec.~\ref{S:masking}. The models for $\Phi$ and $b_I$, and the choice of $m_{\rm th}$ used in this work are detailed in Sec.~\ref{S:models}.

For spectral lines, we build the emission model based on the halo model formalism and the mass $M$ and line luminosity $L_{\rm line}$ relation:
\begin{equation}\label{E:dnuInu_dz_line}
\begin{split}
\left.\frac{d(\nu I_\nu)}{dz}\right|_{\rm line}&(z,\nu)=\int^{M_{\rm th}(z)}_{M_{\rm min}}dM \frac{dn}{dM}(M,z)\frac{\nu L_{\rm line}(M,z)}{4\pi D_L^2(z)}\\
&\cdot\frac{d\chi}{dz}(z)D_A^2(z)\delta^D(\nu-\nu_{\rm rf}^{\rm line}/(1+z)),
\end{split}
\end{equation}
where $dn/dM$ is the halo mass function \citep{2001MNRAS.323....1S},  $D_L$ is the luminosity distance, $\delta^D$ is the Dirac delta function, and $\nu_{\rm rf}^{\rm line}$ is the rest-frame frequency of the spectral line. Here, we treat the intrinsic spectral line width as a delta function, as the intrinsic line width in a galaxy is unresolved in SPHEREx's low-resolution spectral bands. The lower limit of the integration, $M_{\rm min}$, is the minimum halo mass hosting sources with line emission, and we set $M_{\rm min}=10^8 M_\odot$ $h^{-1}$. The upper limit, $M_{\rm th}(z)$, is determined by the line masking threshold and detailed in Sec.~\ref{S:masking}.

The line intensity measured at frequency band $\nu$ with width $\Delta\nu$ is
\begin{equation}\label{E:dnuInu_dz_line_bin}
\left.\frac{d(\nu I_\nu)}{dz}\right|_{\rm line}(z,\nu,\Delta \nu) = \frac{1}{\Delta z^{\rm line}}\left.\nu I_{\nu}\right|_{\rm line}(z^{\rm line}),
\end{equation}
where
\begin{equation}\label{E:nuInu_line}
\begin{split}
\left.\nu I_{\nu}\right|_{\rm line}(z)&=\int^{M_{\rm th}(z)}_{M_{\rm min}}dM \frac{dn}{dM}(M,z)\\
&\cdot\frac{\nu L_{\rm line}(M,z)}{4\pi D_L^2(z)}\frac{d\chi}{d\nu}(z)D_A^2(z),
\end{split}
\end{equation}
$z^{\rm line}=\nu_{\rm rf}^{\rm line}/\nu-1$, $\Delta z^{\rm line}=\Delta\nu\left (\nu_{\rm rf}^{\rm line} /\nu^2  \right )$, and $d\chi/d\nu=c(1+z)^2/(\nu_{\rm rf}^{\rm line} H(z))$. Thus,
\begin{equation}
\begin{split}
\left.\frac{d(\nu I_\nu)}{dz}\right|_{\rm line}(z,\nu, \Delta \nu)  &=\frac{c}{4\pi (1+z)^2 H(z)} \frac{\nu}{\Delta \nu}\\
&\cdot\int^{M_{\rm th}(z)}_{M_{\rm min}}dM \frac{dn}{dM}(M,z)  L_{\rm line}(M,z).\\
\end{split}
\end{equation}
Note the line intensity is inversely proportional to the observed spectral width $\Delta\nu$.

The intensity Poisson noise comprises galaxy continuum and line emission, and foreground emission. Here, we consider Galactic stars as foreground sources that contribute to the fluctuation power spectrum, and thus
\begin{equation}\label{E:ClI_sh}
C_{\ell, \rm P}^I=\left.C_{\ell, \rm P}^I\right|_{\rm cont}+\sum_{\rm line}\left.C_{\ell, \rm P}^I\right|_{\rm line}+\left.C_{\ell, \rm P}^I\right|_{\rm star}.
\end{equation}
In the limit of narrow spectral channel width, the Poisson noise components are given by
\begin{equation}\label{E:ClI_sh_conti}
\begin{split}
\left.C_{\ell, \rm P}^I\right|_{\rm cont}=\int dz &\int_{m_{\rm th}(z,\nu)}^{\infty}dm \Phi(m,z,\nu(1+z))\\
&\cdot\left [\nu F_\nu(m)\right]^2\frac{d\chi}{dz}(z)D_A^2(z),
\end{split}
\end{equation}

\begin{equation}\label{E:ClI_sh_line}
\begin{split}
&\left.C_{\ell, \rm P}^I\right|_{\rm line}=\sum_{\rm line} \frac{1}{\Delta z^{\rm line}}\frac{H(z^{\rm line})}{c\chi^2(z^{\rm line})}
\int^{M_{\rm th}(z)}_{M_{\rm min}}dM\\
&\cdot\frac{dn}{dM}(M,z^{\rm line})\left [\frac{\nu L_{\rm line}(M,z^{\rm line})}{4\pi D_L^2(z^{\rm line})}\frac{d\chi}{d\nu}(z^{\rm line})D_A^2(z^{\rm line})\right]^2,
\end{split}
\end{equation}
and
\begin{equation}\label{E:Cl_sh_star}
\left.C_{\ell, \rm P}^I\right|_{\rm star}=\int_{m_{\rm th}(z,\nu)}^{\infty}dm \frac{dN_{\rm star}}{dmd\Omega}(m,\nu)\left [\nu F_\nu(m)\right]^2,
\end{equation}
where $dN_{\rm star}/dm/d\Omega$ is the star count per magnitude per solid angle. 

Here, we ignore the correlation between continuum and line emissions and model them as independent source populations that contribute to the Poisson noise, while in reality the continuum and line emissions can come from the same or nearby sources that are spatially correlated. The impact of this correlation is negligible in the narrowband measurements ($\Delta\nu/\nu\ll 1$) considered in this work, as the line emission in a given frequency channel only comes from a narrow redshift range ($\Delta z^{\rm line}\ll 1$).

We assume instrument noise to be Gaussian with a variance $\sigma_n^2(\nu)$, so the noise power spectrum is given by
\begin{equation}\label{E:Cln}
C_{\ell,n}^{\rm I}=\sigma_n^2(\nu)\Omega_{\rm pix},
\end{equation}
where $\Omega_{\rm pix}$ is the pixel size. Note that $\sigma_n^2(\nu)$ has the same units as $\nu I_\nu$.

\def\Clx{$C_\ell^x$}
\subsection{Galaxy-intensity Cross Power Spectrum \Clx}\label{S:Clx}
The cross power spectrum also contains clustering and Poisson noise components:
\begin{equation}\label{E:Clx}
C_\ell^{\rm x} = C_{\ell, \rm clus}^{\rm x}+C_{\ell, \rm P}^{\rm x}.
\end{equation}
The clustering term in the Limber approximation has the expression:
\begin{equation}\label{E:Clx_clus}
\begin{split}
C_{\ell, \rm clus}^{\rm x}&=\int dz \frac{H(z)}{c} \frac{f^g(z)f^I(z)}{\chi^2(z)}r_{\rm x}(z)b_g(z)b_I(z)\\
&\cdot \frac{d(\nu I_\nu)}{dz}(z,\nu)P_{\rm m}(k=\frac{\ell+\frac{1}{2}}{\chi(z)},z),
\end{split}
\end{equation}
where $r_{\rm x}$ is the cross-correlation coefficient between a galaxy sample and intensity field. We set $r_{\rm x}=1$ throughout this work as we consider linear scales only.

We separate the clustering term into continuum and line emission,
\begin{equation}
C_{\ell, \rm clus}^{\rm x}=\left.C_{\ell, \rm clus}^{\rm x}\right|_{\rm cont}+\sum_{\rm line}\left.C_{\ell, \rm clus}^{\rm x}\right|_{\rm line}.
\end{equation}
In the limit of a narrow observed spectral channel ($\Delta\nu/\nu\ll 1$) and redshift width ($\Delta z^g\ll 1$), we get
\begin{equation}
\begin{split}
\left.C_{\ell, \rm clus}^{\rm x}\right|_{\rm cont}&=\frac{H(z^g)}{c\chi^2(z^g)}b_g(z^g)b_{I,\rm cont}(z^g)\\
&\cdot \left.\frac{d(\nu I_\nu)}{dz}\right|_{\rm cont}(z^g,\nu)
P_{\rm m}\left(k=\frac{\ell+\frac{1}{2}}{\chi(z^g)},z^g\right),
\end{split}
\end{equation}

\begin{equation}
\begin{split}
\left.C_{\ell, \rm clus}^{\rm x}\right|_{\rm line}&=\frac{H(z^g)}{c\chi^2(z^g)}\frac{\Delta z^{\rm g, line}}{\Delta z^g \Delta z^{\rm line}}b_g(z^g)b_{I,\rm line}(z^g)\\
&\cdot \left.\nu I_{\nu}\right|_{\rm line}(z^g)P_{\rm m}\left(k=\frac{\ell+\frac{1}{2}}{\chi(z^g)},z^g\right),
\end{split}
\end{equation}
where $\Delta z^{\rm g, line}$ is the redshift range where the galaxy tracer redshift $z^g$ overlaps with the line redshift $z^{\rm line}$.

The Poisson cross power spectrum is proportional to the total intensity of the tracer galaxies; in real space, this is nearly equivalent to stacking the intensity at the position of tracer galaxies, or the conditional correlation function. The Poisson noise cross power spectrum can similarly be written as the sum of continuum and line contributions:
\begin{equation}\label{E:Clx_sh}
\begin{split}
C_{\ell, \rm P}^{\rm x}&=\left ( \frac{dN_g}{d\Omega} \right )^{-1}\Delta z^g \left.\frac{d(\nu I_\nu)}{dz}\right|_{g}(z^g,\nu)\\
&=\left.C_{\ell, \rm P}^{\rm x}\right|_{\rm cont}+\left.C_{\ell, \rm P}^{\rm x}\right|_{\rm line}.
\end{split}
\end{equation}
We use ``$g$'' to represent the intensity of the tracer (galaxy) sample. The continuum and line Poisson noise components are given by
\begin{equation}
\left.C_{\ell, \rm P}^{\rm x}\right|_{\rm cont}=\left (\frac{dN_g}{d\Omega} \right )^{-1}\Delta z^g \left.\frac{d(\nu I_\nu)}{dz}\right|_{{\rm cont},g}(z^g,\nu),
\end{equation}
and
\begin{equation}
\left.C_{\ell, \rm P}^{\rm x}\right|_{\rm line}=\left (\frac{dN_g}{d\Omega} \right )^{-1}\frac{\Delta z^{g,\rm line}}{\Delta z^{\rm line}} \left.\nu I_{\nu}\right|_{{\rm line},g}(z^g,\nu),
\end{equation}
where the last term in both equations is the integration of Eq.~\ref{E:dnuInudz_conti} and \ref{E:nuInu_line} over the galaxy sample, respectively. Here, we assume galaxy tracers are the brightest continuum and line emitting sources, and thus the tracer population in our model corresponds to all of the galaxies above a limiting magnitude $m_g(z,\nu)$ and its corresponding halo mass $M_g(z)$. For a given masking threshold $m_{\rm th}(z,\nu)$; (corresponding to halo mass threshold $M_{\rm th}(z)$), if the tracer catalog is deeper than the masking threshold, i.e., $m_g > m_{\rm th}$ ($M_g < M_{\rm th}$), we get
\begin{equation}
\begin{split}
\left.\frac{d(\nu I_\nu)}{dz}\right|_{{\rm cont},g}(z,\nu)=\int_{m_{\rm th}(z,\nu)}^{m_g(z,\nu)}&dm \Phi(m,z,\nu(1+z))\\
\cdot\nu F_\nu(m)&\frac{d\chi}{dz}(z)D_A^2(z),
\end{split}
\end{equation}
and
\begin{equation}
\begin{split}
\left.\nu I_{\nu}\right|_{{\rm line},g}(z,\nu)=\int^{M_{\rm th}(z)}_{M_g(z)} dM& \frac{dn}{dM}(M,z)\\
\cdot\frac{\nu L_{\rm line}(M,z)}{4\pi D_L^2(z)}&\frac{d\chi}{d\nu}(z)D_A^2(z).
\end{split}
\end{equation}
By contrast, if $m_g \leqslant m_{\rm th}$ ($M_g \geqslant M_{\rm th}$), all tracer galaxies are masked, and thus
\begin{equation}\label{E:dnuInu_dz_cont_g_0}
\left.\frac{d(\nu I_\nu)}{dz}\right|_{{\rm cont},g}(z,\nu)=\left.\nu I_{\nu}\right|_{{\rm line},g}(z,\nu)=0,
\end{equation}
and the cross Poisson power vanishes.

\def\Txx{$\mathcal{T}^{gIgI}$}
\subsection{Galaxy-intensity Cross Trispectrum \Txx}\label{S:Txx}
We model the non-Gaussian contribution in the Poisson regime using the trispectrum between the galaxy and intensity fields, given by  \citet{2018PhRvD..97l3539S}
\begin{equation}
\mathcal{T}^{gIgI}(\ell,\ell') = \left [\left.C_{\ell, \rm P}^I\right|_{{\rm cont},g}+\sum_{\rm line}\left.C_{\ell, \rm P}^I\right|_{{\rm line},g}  \right ] \cdot \left (C_{\ell, \rm P}^g  \right )^2,
\end{equation}
where $C_{\ell, \rm P}^g$ is described by Eq.~\ref{E:Clg_shot},  and $\left.C_{\ell, \rm P}^I\right|_{\rm cont,g}$ and $\left.C_{\ell, \rm P}^I\right|_{\rm line,g}$ are the integration of Eq.~(\ref{E:ClI_sh_conti}) and Eq.~(\ref{E:ClI_sh_line}) over the galaxy sample, respectively. In the Poisson regime, the $\mathcal{T}^{gIgI}(\ell,\ell')$ function is a constant for all values of $\ell$ and $\ell'$ pairs, and specifies the off-diagonal terms of the covariance matrix (Eq.~\ref{E:Cov}) that describes the correlation between different multipole modes.
If $m_g > m_{\rm th}$ ($M_g < M_{\rm th}$),
\begin{equation}
\begin{split}
\left.C_{\ell, \rm P}^I\right|_{{\rm cont},g}=\int dz \int_{m_{\rm th}(z,\nu)}^{m_g(z,\nu)}&dm \Phi(m,z,\nu(1+z))\\
&\cdot\left [\nu F_\nu(m)\right]^2\frac{d\chi}{dz}(z)D_A^2(z)
\end{split}
\end{equation}
and
\begin{equation}
\begin{split}
\left.C_{\ell, \rm P}^I\right|_{{\rm line},g}&=\sum_{\rm line} \frac{1}{\Delta z^{\rm line}}\frac{H(z^{\rm line})}{c\chi^2(z^{\rm line})}\\
&\cdot\int^{M_{\rm th}(z)}_{M_g(z)}dM\cdot\frac{dn}{dM}(M,z^{\rm line})\\
&\cdot\left [\frac{\nu L_{\rm line}(M,z^{\rm line})}{4\pi D_L^2(z^{\rm line})}\frac{d\chi}{d\nu}(z^{\rm line})D_A^2(z^{\rm line})\right]^2.
\end{split}
\end{equation}
Otherwise, if $m_g \geqslant m_{\rm th}$ ($M_g \leqslant M_{\rm th}$), we get
\begin{equation}
\left.C_{\ell, \rm P}^I\right|_{{\rm cont},g} = \left.C_{\ell, \rm P}^I\right|_{{\rm line},g} = \mathcal{T}^{gIgI}=0.
\end{equation}

\section{Emission Model}\label{S:models}
We assume the integrated galaxy light (IGL) constitutes the majority of the near-infrared emission, and model the spectral energy distribution as continuum and spectral line components. In addition, we consider a model for quasars to study constraints on quasar emission in the Poisson regime. Note we do not add the quasar model to the IGL model, since IGL, as described below, already includes the quasar contribution. We also consider Galactic stars as a foreground emission.

\subsection{Galaxy Spectral Energy Distribution}
\subsubsection{Continuum Emission}\label{S:conti_emission}
Our model for the IGL continuum emission is based on a luminosity function prescription from \cite{2012ApJ...752..113H}, where the luminosity function, $\Phi(m,z,\nu_{\rm rf})$, has a Schechter functional form \citep{1976ApJ...203..297S} that depends on magnitude $m$, redshift $z$, and rest-frame frequency $\nu_{\rm rf}$. The parameters are calibrated to observations in several ultraviolet to mid-infrared wavelength bands, and we interpolate the Schechter parameters to all wavelengths from their model values.

Our model for the continuum intensity bias, $b_{I, \rm cont}(z,\nu)$, is also from \cite{2012ApJ...752..113H}, who uses a halo occupation distribution (HOD) framework \citep{2005ApJ...633..791Z} and assumes no frequency dependence:
\begin{equation}\label{E:bI_conti}
b_{I, \rm cont}(z)=\frac{\int dM \frac{dn}{dM}(M,z) b(M,z) \left \langle N_{\rm gal} \right \rangle}{\int dM \frac{dn}{dM}(M,z) \left \langle N_{\rm gal} \right \rangle},
\end{equation}
where $\left \langle N_{\rm gal} \right \rangle$ is the total halo occupation number. Following \cite{2012ApJ...752..113H}, we use the HOD parameters given by the Sloan Digital Sky Survey (SDSS) measurement in \cite{2011ApJ...736...59Z}.

\subsubsection{Line Emission}
In this work, we consider five prominent optical spectral lines: H$\alpha$ (656.3 nm), [\ion{O}{iii}] (500.7 nm), H$\beta$ (486.1 nm), [\ion{O}{ii}] (372.7 nm), and Ly$\alpha$ (121.6 nm). These lines are of particular interest as they are visible in SPHEREx bands across a range of redshifts.

Our line emission model is built on an empirical halo mass and line luminosity relation $L_{\rm line}(M,z)$. We adopt and summarize the prescription from \cite{2017ApJ...835..273G} below. First, we assume a simple linear relation between star formation rate (SFR) and halo mass, and calibrate the scaling factor using the star formation rate density constraints from \cite{2006ApJ...651..142H}. We then further assume a linear relation between the line luminosity and SFR, using SFR--$L_{\rm line}$ relations from \citet{1998ARA&A..36..189K} and \citet{2007ApJ...657..738L} for H$\alpha$, [\ion{O}{ii}], and [\ion{O}{iii}] emission:
\begin{align}
\frac{\rm SFR}{M_\odot/ \rm{yr}} &= 7.9\times 10^{-42} \frac{L_{H\alpha}}{\rm{erg / s}},\\
\frac{\rm SFR}{M_\odot/ \rm{yr}} &= 1.4 \times 10^{-41} \frac{L_{[O\,II]}}{\rm{erg / s}},\\
\frac{\rm SFR}{M_\odot/ \rm{yr}} &= 7.6 \times 10^{-42} \frac{L_{[O\,III]}}{\rm{erg / s}}.
\end{align}
For H$\beta$, we assume a fixed line ratio of H$\beta$/H$\alpha=0.35$ \citep{2006agna.book.....O}. As Ly$\alpha$ emission is not discussed in \cite{2017ApJ...835..273G}, we use a model with a constant halo mass to luminosity M--$L_{{\rm Ly}\alpha}$ ratio at halo masses of $M = 10^8$--$10^{15}$ Mpc $h^{-1}$, and calibrate the scaling factor to match the mean Ly$\alpha$ intensity of the analytical model from \citet[][Eq. 51]{2014ApJ...786..111P}.

We note that the emission line strength highly depends on individual galaxy properties, so in reality, there will be a large scatter in the 
halo mass--line luminosity relation. Nevertheless, we check that the mean line luminosity density from our model is consistent with various observations \citep{2010ApJ...714..255G,2012ApJ...744..110C,2013MNRAS.428.1128S,2013MNRAS.431.3589Z,2015MNRAS.452.3948K} compiled in \citet{2017MNRAS.464.1948F}. As we are probing the averaged emission with cross-correlation, our simple line emission model is sufficient for this work, and we leave more detailed models to future work.

The line emission bias factor, $b_{I, \rm line}$, is given by the the halo bias, $b(M,z)$, weighted by line luminosity:
\begin{equation}
b_{I, \rm line}(z)=\frac{\int_{M_{\rm min}}^{M_{\rm th}(z)} dM \frac{dn}{dM}(M,z) b(M,z)L_{\rm line}(M,z)}{\int_{M_{\rm min}}^{M_{\rm th}(z)} dM \frac{dn}{dM}(M,z)L_{\rm line}(M,z)}.
\end{equation}
We use the same lower and upper integration limits ($M_{\rm min}$ and $M_{\rm th}(z)$) as Eq.~\ref{E:dnuInu_dz_line}.

\subsection{Quasar Spectrum}\label{S:QSO}
Our quasar emission model assumes a quasar luminosity function and an averaged quasar spectrum. We adopt the fiducial model of the quasar luminosity function in \cite{2013ApJ...762...70C}, and use the quasar spectrum template compiled in \cite{2018AJ....156...66M} that splices together a composite spectra from \cite{2001AJ....122..549V}, \cite{2006ApJS..166..470R}, and \cite{2015MNRAS.449.4204L}.

\subsection{Foreground Stars}
Galactic stars contribute to the Poisson noise (Eq.~\ref{E:Cl_sh_star}) in the auto spectrum of the intensity field. We model star counts, $dN_{\rm star}/dm/d\Omega$ in Eq.~\ref{E:Cl_sh_star}, using the stellar population synthesis code, Trilegal\footnote{\url{http://stev.oapd.inaf.it/cgi-bin/trilegal}} \citep{2002A&A...391..195G,2005A&A...436..895G}. We use their model in the \textit{UBVRIJHKLMN} photometric system at the North Galactic Pole ($\ell=0^\circ$, $b=90^\circ$).

\section{Cross-correlation Forecast}\label{S:xcorr_forecast}
\subsection{Evaluation Metrics}\label{S:metrics}
We focus on constraining the  redshift-dependent IGL spectrum, $d(\nu I_\nu)/dz(\nu,z)$, from cross-correlating galaxies and the intensity field. The clustering and Poisson noise amplitudes in the cross power spectrum are proportional to $b_I(z)d(\nu I_\nu)/dz(z,\nu)$ and $\left.d(\nu I_\nu)/dz\right|_{\rm g}(z,\nu)$, respectively (Equations .~\ref{E:Clx_clus} and \ref{E:Clx_sh}). Therefore we forecast the signal-to-noise ratio (SNR) on these two quantities obtained from the cross power spectrum. Note $\left.d(\nu I_\nu)/dz\right|_{\rm g}(z,\nu)$ represents the IGL spectrum from the specific galaxies (tracers) used in the cross-correlation measurements, and does not equal the cosmic mean IGL spectrum, $d(\nu I_\nu)/dz(z,\nu)$.

As the matter power spectrum is linear on large scales, here defined as $k\lesssim 0.2$ $h$ Mpc$^{-1}$, we evaluate constraints on $b_I(z)d(\nu I_\nu)/dz(z,\nu)$ from the amplitude of all $\ell_{\rm clus}^{\rm min}<\ell<\ell_{\rm clus}^{\rm max}$ modes, where $\ell_{\rm clus}^{\rm max}=0.2{\rm [h/Mpc]}\chi(z)-1/2$, and the minimum accessible mode $\ell_{\rm clus}^{\rm min}=51$, which corresponds to SPHEREx's field of view of $3.5^{\circ}$ on the smaller side, as we conservatively assume modes larger than the field size will be susceptible to the zodiacal light foreground filtering process and other large-scale systematics and become inaccessible. The SNR on the clustering amplitude is
\begin{equation}\label{E:SNR_clus}
{\rm SNR}_{\rm clus}(b_I(z)\frac{d(\nu I_\nu)}{dz}(z,\nu))=\sqrt{\sum_{\ell \in \left [ \ell_{\rm clus}^{\rm min}, \ell_{\rm clus}^{\rm max} \right ]}\left [ \frac{C^{\rm x}_{\ell, \rm clus}}{\delta C_\ell^{\rm x}} \right ]^2}.
\end{equation}

In the Poisson regime, defined as $k\gtrsim 1$ $h$ Mpc$^{-1}$, we use $\ell$ modes in the range of $\ell_{\rm P}^{\rm min}<\ell<\ell_{\rm P}^{\rm max}$ to constrain $\left.\frac{d(\nu I_\nu)}{dz}\right|_{\rm g}(z,\nu)$,  where $\ell_{\rm P}^{\rm min}=1{\rm [h/Mpc]}\chi(z)-1/2$, and $\ell_{\rm P}^{\rm max}=10^5$, which is the highest multipole mode available given a SPHEREx pixel size of $6''.2$. Incorporating the effects from both Gaussian and non-Gaussian terms, the SNR on the Poisson amplitude is given by
\begin{equation}\label{E:SNR_shot}
\begin{split}
&{\rm SNR}_{\rm P}(\left.\frac{d(\nu I_\nu)}{dz}\right|_g(z,\nu))\\
&= \frac{C^{\rm x}_{\ell, \rm P}} {\sqrt{\frac{1}{N_{\ell>\ell_{\rm min}}}\sum_{\ell \in \left [ \ell_{\rm P}^{\rm min}, \ell_{\rm P}^{\rm max} \right ]}\left ( \delta C_\ell^{\rm x} \right )^2 + \frac{\mathcal{T}^{gIgI}}{\Omega_{\rm sur}}}},
\end{split}
\end{equation}
where $\Omega_{\rm sur}$ is the survey area solid angle used for cross-correlation.

Since $b_I(z)d(\nu I_\nu)/dz(z,\nu)$ and $\left.d(\nu I_\nu)/dz\right|_g(z,\nu)$ are the main quantities of interest, we forecast below their constraints as a function of redshift and wavelength from upcoming surveys.

\subsection{Masking}
In practice, pixels that contain bright stars and bright galaxies will have to be masked in data processing to reduce sample variance noise in the cross power spectrum, and the resulting power spectrum amplitude will depend on the depth of the masking limit. To model the effect of masking on the power spectrum, we assume a photometric source catalog with sufficient depth provides the location and fluxes of bright stars and galaxies to be masked. The masking magnitude threshold, $m_{\rm th}^{\rm ph}$, is set by the limiting magnitude of a photometric band at frequency $\nu^{\rm ph}$ (or the corresponding wavelength $\lambda^{\rm ph}$).

For simplicity, we mask all stars and galaxies brighter than our masking threshold (see Sec.~\ref{S:masking} for discussion). Since only the flux density information is needed, we can use a photometric source catalog for this purpose, which is usually deeper than a spectroscopic catalog. Given $m_{\rm th}^{\rm ph}$ and $\nu^{\rm ph}$, we adopt the abundance matching concept \citep{2004ApJ...609...35K,2004MNRAS.353..189V,2006ApJ...647..201C,2010ApJ...717..379B,2010ApJ...710..903M} to infer the masking magnitude threshold $m_{\rm th}(\nu,z)$ for continuum emission at a given frequency $\nu$; this assumes that the brightest $N$ sources in a given frequency bin are also the brightest ones in another frequency bin. We therefore first calculate the number density of galaxies being masked at $\nu^{\rm ph}$:
\begin{equation}
N = \int^{m_{\rm th}^{\rm ph}}_{-\infty} dm \Phi(m,z,\nu^{\rm ph}(1+z)),
\end{equation}
and then at each frequency solve for the $m_{\rm th}(\nu,z)$ value that gives
\begin{equation}
N = \int^{m_{\rm th}(\nu,z)}_{-\infty} dm \Phi(m,z,\nu(1+z)).
\end{equation}
Analogously, for line emission modeled with a line luminosity and halo mass relation, given the halo mass function we can derive its maximum halo mass $M_{\rm th}(z)$ by solving for 
\begin{equation}
N=\int^{\infty}_{M_{\rm th}(z)}dM \frac{dn}{dM}(M,z).
\end{equation}

Throughout this work, we set the fiducial masking threshold to $m_{\rm th}^{\rm ph}=20$ at frequency $\nu^{\rm ph}=3 \times 10^{14}$ Hz (1 $\mu$m). Our model suggests that $\sim 3\%$ of the SPHEREx pixel ($6''.2$) contains stars or galaxies above this limit, and the intensity Poisson noise from unmasked stars is approximately an order of magnitude below the galaxy Poisson noise. For simplicity we ignore the sensitivity loss from mode mixing due to masking, and discuss the implications in Sec.~\ref{S:masking}.

\subsection{Fiducial Case}\label{S:fiducial_case}
We consider cross power spectra between the SPHEREx intensity map and several current and future galaxy redshift surveys. For simplicity, we present results with a few chosen parameters as our fiducial case.

\begin{figure}[ht!]
\begin{center}
\includegraphics[width=\linewidth]{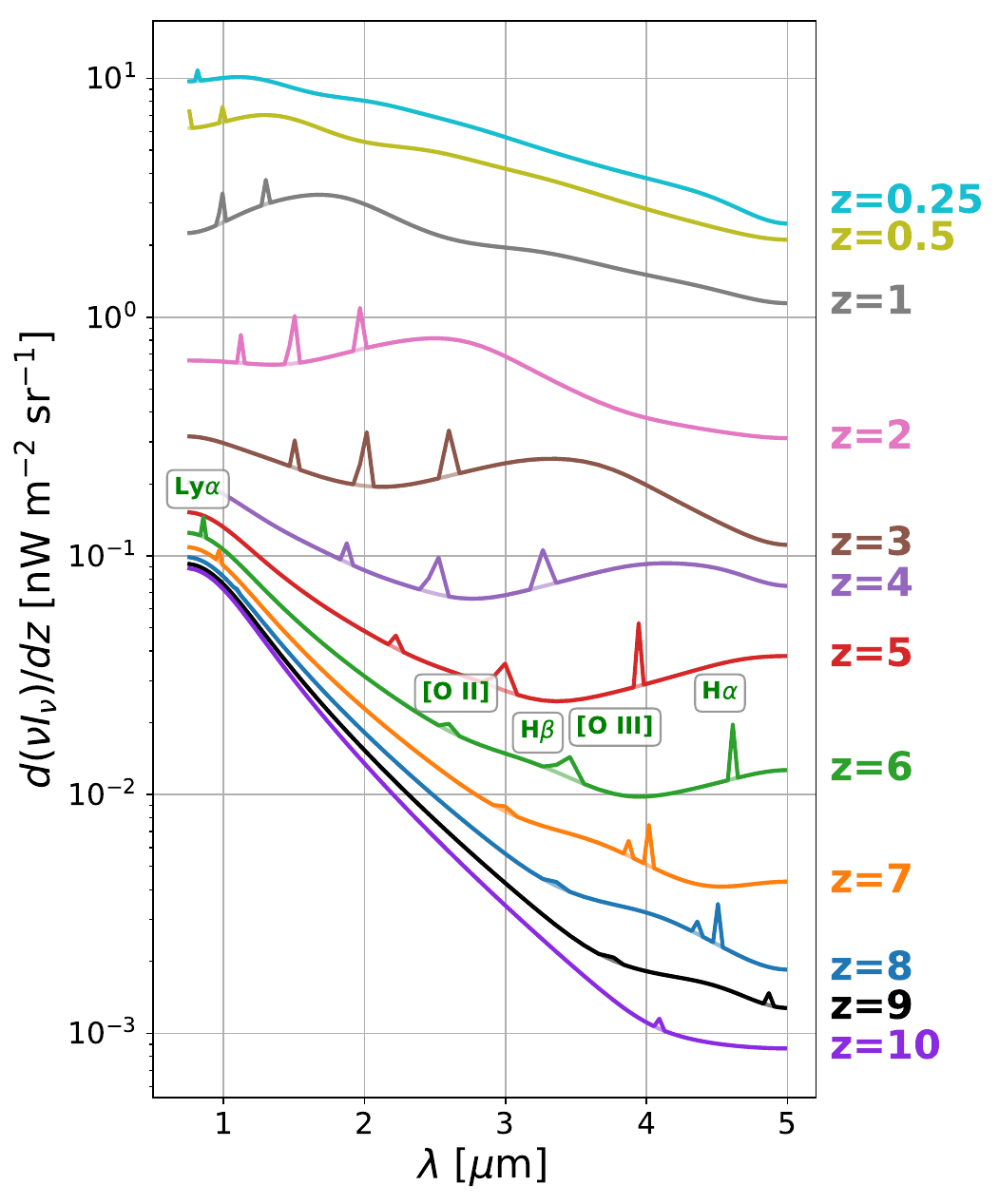}
\caption{\label{F:nuInu} Redshift-dependent IGL spectra from our model (Sec.~\ref{S:models}) at fiducial redshifts, using SPHEREx spectral binning. The spectra contain continuum and line emission. The spectral lines from left to right are Ly$\alpha$, [\ion{O}{ii}], H$\beta$, [\ion{O}{iii}], and H$\alpha$, respectively, as labeled in the $z=6$ case (in green). At $z<7$, H$\beta$ and [\ion{O}{iii}] lines are in adjacent spectral channels and thus appear as one blended line feature; whereas at $z\geq 7$, the two lines are redshifted into SPHEREx's high spectral resolution bands (see Fig.~\ref{F:SPHEREx_noise}), which split the lines by several channels, and thus two peaks are distinguishable in the plot. Note here we present our model on infinitesimally thin redshift slices, and ignore the intrinsic line width and broadening effects of all spectral lines, so the line width shown only reflects the impact of the SPHEREx spectral resolution.}
\end{center}
\end{figure}

Fig.~\ref{F:nuInu} shows the redshift-dependent IGL spectra, $d(\nu I_\nu)/dz$, from our model (Sec.~\ref{S:models}) at fiducial redshifts defined below. The spectra include continuum and spectral line emission and are presented with SPHEREx spectral binning (Sec.~\ref{S:SPHEREx}) of 96 channels over $0.75$--$5$ $\mu$m.\footnote{SPHEREx recently modified its design to have 102 frequency channels covering the same wavelength range. As this has a negligible impact on our results, we use their 96 channel configuration with published sensitivity forecasts for this work.} The IGL spectral amplitude builds up with cosmic time. Since SPHEREx has higher spectral resolution at $\lambda>3.8$ $\mu$m (see Sec.~\ref{S:SPHEREx}), the line intensity at longer wavelengths is stronger compared to the continuum as the line signal is inversely proportional to the spectral width (Eq.~\ref{E:dnuInu_dz_line_bin}). The model we used for the IGL continuum \citep{2012ApJ...752..113H} is only calibrated to observations at $z\lesssim 5$ and rest-frame wavelengths $\lambda > 0.15$ $\mu$m. The higher redshifts and shorter wavelengths models are thus derived from extrapolation.

At low redshifts ($z<3$), cross-correlations between SPHEREx and spectroscopic redshift surveys with high redshift accuracy ($\sigma_z/(1+z)\sim 10^{-4}$--$10^{-3}$) allow for the selecting of galaxies in thin redshift slices. We choose a redshift bin width of $\Delta z^g/(1+z^g)$=0.03, and calculate the cross-correlation at five fiducial redshifts, $z^g=[0.25, 0.5, 1, 2, 3]$. Between a redshift of $0$ and $3$ there are $\sim 46$ such measurements. We choose a fiducial masking threshold $m_{\rm th}^{\rm ph}=20$ at $\lambda^{\rm ph}=1$ $\mu$m for all of these cases, except for at $z=0.25$ where we use $m_{\rm th}^{\rm ph}=18$. This threshold is chosen to optimize the trade-off between losing signal and reducing foreground emission (see Sec.~\ref{S:masking} for more discussion). SPHEREx will achieve an all-sky point-source sensitivity of $m_{\rm AB}\sim 20$ at $\lambda/\Delta\lambda\sim 4$, and therefore our fiducial masking depth is feasible with the SPHEREx source catalog.

At higher redshifts ($z>3$), spectroscopic catalogs are either unavailable or lacking sufficient number density and/or sky coverage, and therefore we consider cross-correlating SPHEREx with photometric redshift catalogs, which have a redshift accuracy of  $\sigma_z/(1+z)\sim0.02$--$0.05$. We choose eight fiducial redshifts for the calculation: $z^g=[3, 4, 5, 6, 7, 8, 9, 10]$. At $3\leqslant z \leqslant 6$, we use a redshift bin width of $\Delta z^g/(1+z^g)=0.1$, which results in $\sim6$ such measurements. At $z>6$, the redshift bins are determined by the expected redshift resolution from  the Lyman-break dropout technique (see Sec.~\ref{S:phot_surveys} for details). We also set the masking threshold to $m_{\rm th}^{\rm ph}=20$ at $\lambda^{\rm ph}=1$ $\mu$m.

\section{SPHEREx Intensity Mapping}\label{S:SPHEREx}
\begin{figure}[ht!]
\includegraphics[width=\linewidth]{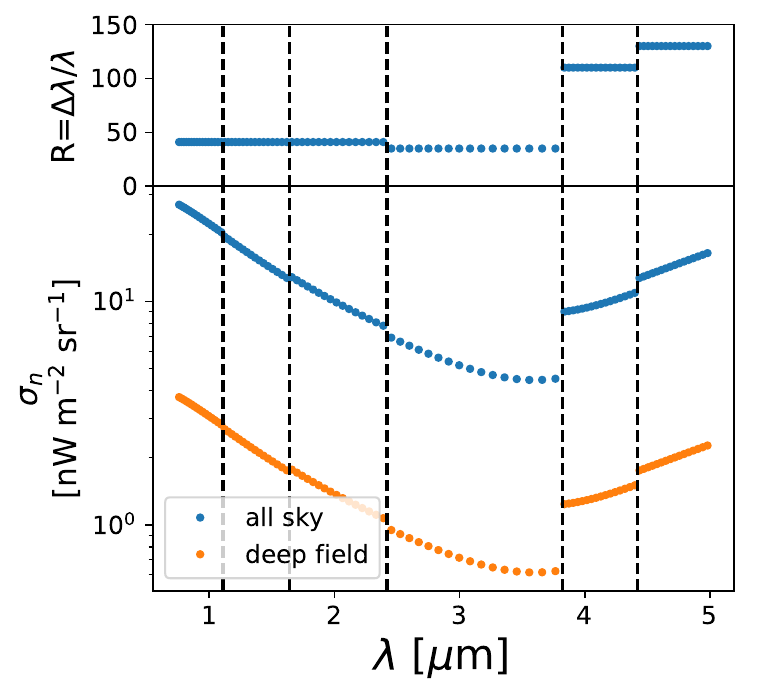}
\caption{\label{F:SPHEREx_noise}Top: SPHEREx spectral resolution of each wavelength channel. The vertical dashed lines mark the six wavelength bands, each containing 16 channels. Bottom: SPHEREx surface brightness sensitivity per spectral channel for all sky (blue) and  deep field (orange) in a $6''.2$ sky pixel.}
\end{figure}

SPHEREx is a NASA MIDEX mission scheduled to launch in 2024 \citep{2014arXiv1412.4872D,2018arXiv180505489D}\footnote{\url{http://spherex.caltech.edu}}. SPHEREx will carry out the first all-sky near-infrared spectro-imaging survey from 0.75 to 5 $\mu$m. It has a pixel size of $6''.2$, and a wavelength-dependent spectral resolution: a spectral resolving power of $R=41$ at wavelengths between 0.75 and 2.42 $\mu$m, (48 spectral channels), $R=35$ between 2.42 and 3.82 $\mu$m (16 spectral channels), $R=110$ between 3.82 and 4.42 $\mu$m (16 spectral channels), and $R=130$ between 4.42 and 5.00 $\mu$m (16 spectral channels), The top panel of Fig.~\ref{F:SPHEREx_noise} shows the SPHEREx spectral resolution as a function of wavelength (spectral channel), and the bottom panel is the expected noise rms per spectral channel per pixel for the all-sky and the two $\sim 100$ deg$^2$ deep surveys. Here, we adopt the expected SPHEREx surface brightness noise levels, which are publicly available\footnote{Data downloaded from: \url{https://github.com/SPHEREx/Public-products/blob/master/Surface_Brightness_v28_base_cbe.txt}}, and calculate the noise contribution to the cross power spectrum (Eq.~\ref{E:Cln}). We use the optimistic sensitivity model, noting that the pessimistic forecast is $\sim 0.5$ mag lower. For the SPHEREx all-sky survey, we only use 75\% of the full sky coverage to avoid Galactic contamination.

\section{Redshift Surveys}\label{S:data}
\begin{figure}[ht!]
\begin{center}
\includegraphics[width=\linewidth]{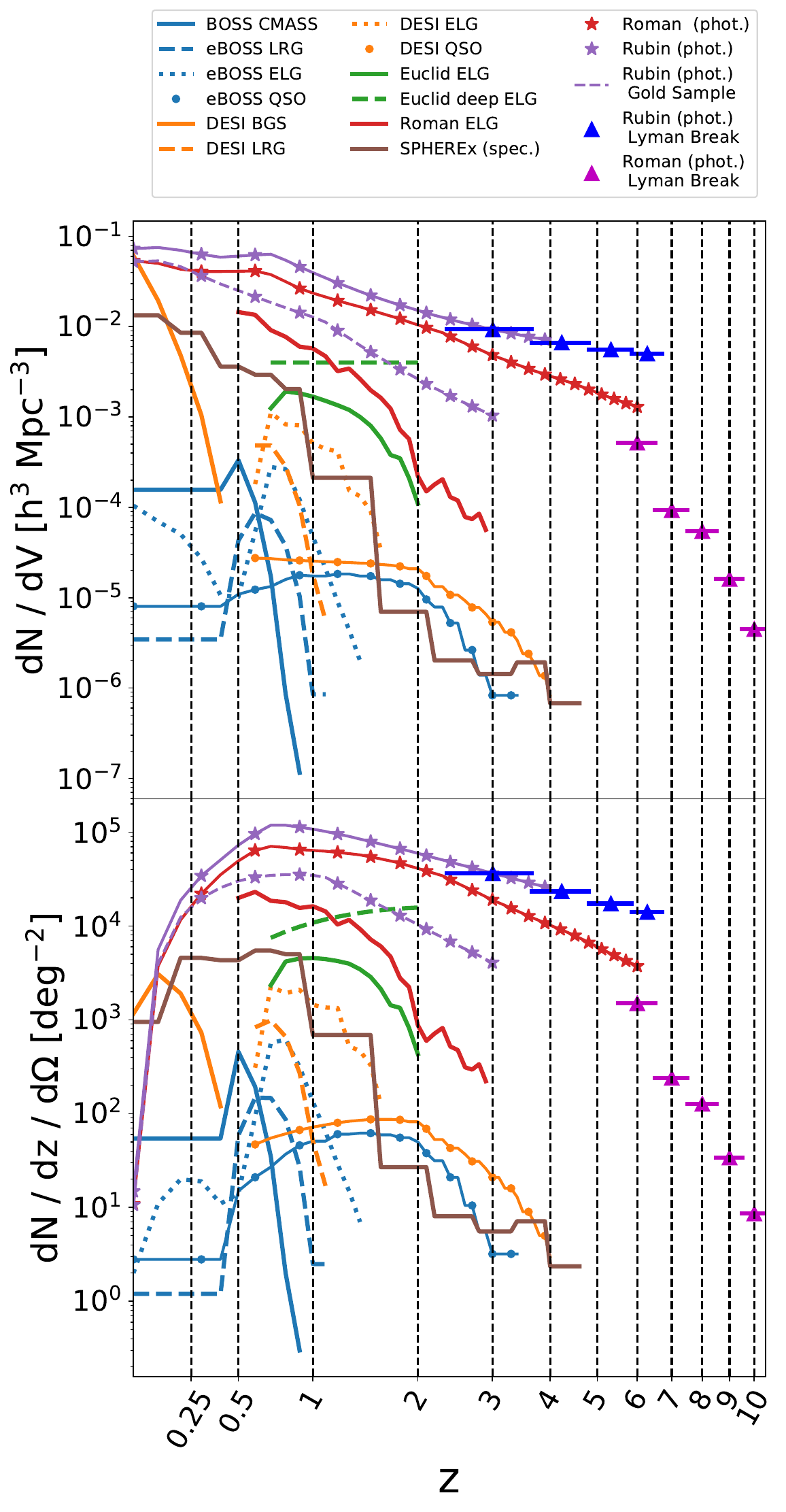}
\caption{\label{F:survey_number_density} Model of number density of the spectroscopic and photometric galaxy catalogs from each survey considered in this work. The values are compiled from the literature or predicted with our model, as detailed in Sec.~\ref{S:data}. We show the number density per co-moving volume in the top panel, and the number density per redshift and solid angle in the bottom panel. The vertical dashed lines mark the fiducial redshifts used in this work.}
\end{center}
\end{figure}

We consider cross-correlating intensity maps from SPHEREx and several current and future galaxy surveys. Table~\ref{T:surveys} summarizes the survey parameters of all galaxy surveys. The model for source number density as a function of redshift is shown in Fig.~\ref{F:survey_number_density}.

\begin{deluxetable}{cccccc}[h]
\tablenum{1}
\tablecaption{\label{T:surveys}Summary of the Sky Area and Redshift Coverage of the Surveys Used in This Work}
\tablewidth{0pt}
\tablehead{
\colhead{Survey Name} & \colhead{$A_{\rm sur}$ [deg$^2$]}  & \colhead{$f_{\rm sky}$} &  \colhead{Redshift Range}
}
\startdata
\hline\hline
Intensity Maps\\\hline\hline
SPHEREx  &30,940    & 0.75   &N/A\\ 
SPHEREx deep  &2$\times$100       &$4.85\times 10^{-3}$ &N/A\\\hline\hline
Spectroscopic Surveys\\\hline\hline
BOSS CMASS    &10,000    &0.242 &$0<z<1$\\
eBOSS LRG     &7500     &0.182 &$0<z<1.2$\\
eBOSS ELG     &620       &0.0150 &$0<z<1.5$\\
eBOSS QSO     &7500     &0.182 &$0<z<3.5$\\\hline
DESI BGS      &14,000    &0.339 &$0<z<0.5$\\
DESI LRG      &14,000    &0.339 &$0.6<z<1.2$\\
DESI ELG      &14,000    &0.339 &$0.6<z<1.7$\\
DESI QSO      &14,000    &0.339 &$0.6<z<4.2$\\\hline
Euclid ELG    &15,000    &0.364 &$0.65<z<2.05$\\
Euclid deep ELG & 10   &$2.42\times 10^{-4}$ &$0.65<z<2.05$\\\hline
Roman ELG     &2200     &0.053 &$0.5<z<3$\\\hline
SPHEREx spec.    &30,940     &0.75 &$0<z<4.6$\\\hline\hline
Photometric Surveys\\\hline\hline
Rubin phot.    &18,000     &0.436 &$0<z<4$\\
Rubin phot.\\ (gold sample)    &18,000     &0.436 &$0<z<3$\\\hline
Rubin Lyman-break    &18,000     &0.436 & $2.4<z<3.7$\\
    &18,000     &0.436 & $3.7<z<4.8$\\
   &18,000     &0.436 & $4.8<z<5.9$\\
   &18,000     &0.436 & $5.9<z<6.7$\\\hline
Roman phot.    &2,200     &0.053 &$0<z<6$\\\hline
Roman Lyman-Break    &2,200     &0.053 & $5.5<z<6.5$\\
    &2200     &0.053 & $6.5<z<7.5$\\
   &2200     &0.053 & $7.5<z<8.5$\\
    &2200     &0.053 & $8.5<z<9.5$\\
    &2200     &0.053 & $9.5<z<10.5$\\
\enddata
\end{deluxetable}

\subsection{Spectroscopic Galaxy Surveys}\label{S:spec_surveys}
We consider the following current and planned future spectroscopic redshift surveys: SDSS BOSS/eBOSS, DESI, Euclid, and the High Latitude Survey with the Roman Space Telescope. Each survey has one or multiple types of density tracers. Below we describe the parameters we use for each tracer catalog, including the number density, sky coverage, and the galaxy bias factors.

\subsubsection{SDSS BOSS/eBOSS}
SDSS BOSS \citep{2013AJ....145...10D} and eBOSS \citep{2016AJ....151...44D} are spectroscopic surveys targeting several types of galaxy tracers. In this work, we use four spectroscopic samples from the SDSS: the BOSS CMASS survey, eBOSS LRG (luminous red galaxies), eBOSS ELG (emission line galaxies), and eBOSS QSO (quasars) surveys. 

For BOSS CMASS, eBOSS LRG, and eBOSS QSO, we use the predicted number density from \citet{2016AJ....151...44D} Table 1 ($z_{\rm conf}>1$ for LRG, and \texttt{QSO\_CORE} New + Known for QSO); for eBOSS ELG, the number density is given in Table 4 of \citet{2017MNRAS.471.3955R}. We assume a top-hat functional form of $dN/dV (z)$, the number density per co-moving volume, between the redshift bins quoted in their tables.
Our model for galaxy bias, $b_g$, also follows the prescription of \citet{2016AJ....151...44D} and \citet{2017MNRAS.471.3955R}: $b_g({\rm LRG})=1.7/\sigma_8(0)/\sigma_8(z)$, $b_g({\rm ELG})=1.0/\sigma_8(0)/\sigma_8(z)$, $b_g({\rm QSO})=0.53 + 0.29(1+z)^2$, and we set $b_g({\rm CMASS})=2$ according to \citet{2013MNRAS.432..743N}. 

\subsubsection{DESI}
The Dark Energy Spectroscopic Instrument \citep[DESI; ][]{2016arXiv161100036D} is an ongoing spectroscopic survey that observes four types of tracers across 14,000 deg$^2$:  bright galaxy samples (BGS), luminous red galaxies (LRG), star-forming emission line galaxies (ELG), and quasars (QSO). The predicted number density for each tracer catalog is given in Tables 2.3, 2.5, and 2.7 of \citet{2016arXiv161100036D}, and we assume a top-hat functional form of $dN/dV (z)$ between the redshift bins quoted in their tables. For LRG, ELG, and QSO, we use the same bias model as for the eBOSS: $b_g({\rm LRG})=1.7/\sigma_8(0)/\sigma_8(z)$, $b_g({\rm ELG})=1.0/\sigma_8(0)/\sigma_8(z)$, $b_g({\rm QSO})=0.53 + 0.29(1+z)^2$, and we set $b_g({\rm BGS})=2$, as for the BOSS CMASS case.

\subsubsection{Euclid}
The Euclid spectroscopic survey \citep{2011arXiv1110.3193L} will map ELGs (H$\alpha$ emitters) over 15,000 deg$^2$ between redshifts $0.9\lesssim z \lesssim 2$. The expected number density as a function of redshift is given by Table 3 of \citet[][we use the $n_2$ reference case]{2018LRR....21....2A}, and we assume a top-hat $dN/dV (z)$ between the redshift bins quoted in their table. For galaxy bias, we use the model from \citet{2019MNRAS.486.5737M}: $b_g(z)=0.7z+0.7$.

In addition, Euclid will also have deep fields that overlap with the SPHEREx deep field at the North Ecliptic Pole. We therefore consider a case of cross-correlating the Euclid deep field catalog with SPHEREx deep field images. For Euclid deep field sources, we adopt the same bias and redshift coverage as the wide-field ELGs and assume a constant number density of $n_g=4\times 10^{-3}$ ($h$ Mpc$^{-1}$)$^3$ over this redshift range.

\subsubsection{Roman Space Telescope}
The Roman Space Telescope \citep{2015arXiv150303757S} plans a spectroscopic survey of ELGs using the H$\alpha$ and [\ion{O}{iii}] lines as part of the High Latitude Survey \citep[HLS; ][]{2018arXiv180403628D}. Compared to Euclid, the Roman Space Telescope--HLS covers a smaller sky area (2200 deg$^2$) but with better point-source sensitivity. We adopt the ELG galaxy number density predictions from \citet{2019MNRAS.490.3667Z}. We use the H$\alpha$ number density (Table 1) for $z<2$, the [\ion{O}{iii}] number density (Table 2) for $2<z<3$, and assume a top-hat $dN/dV (z)$ between the redshift bins quoted in their tables. For both spectral lines we use their ``dust model fit at high redshifts'' and the $1\times 10^{-16}$ erg s$^{-1}$ cm$^{-2}$ flux limit case.

\subsubsection{SPHEREx Spectroscopic Catalog}
In addition to diffuse imaging, SPHEREx will also produce a spectroscopic galaxy catalog \citep{2014arXiv1412.4872D,2018arXiv180505489D}. We therefore also consider the case of cross-correlating the SPHEREx galaxy catalog with its intensity maps. For the SPHEREx all-sky survey, we use the predicted number density and galaxy bias of the SPHEREx source catalog at $z<4.6$ from their public products \footnote{\url{https://github.com/SPHEREx/Public-products/blob/master/galaxy_density_v28_base_cbe.txt}}. Specifically, we use the forecast with a galaxy photometric redshift error of $0.01<\sigma_z/(1+z)<0.03$, and consider a tophat functional form of $dN/dV(z)$ between the quoted redshift bins. The SPHEREx deep field with higher sensitivity will produce a deeper source catalog, and we leave its application to future work.

\subsection{Photometric Galaxy Surveys}\label{S:phot_surveys}
We consider cross-correlating SPHEREx and two upcoming photometric surveys: the Rubin Observatory Legacy Survey of Space and Time (LSST) and the Roman Space Telescope High Latitude Survey (HLS). Despite the larger redshift uncertainties in photometric samples, they have the advantage of reaching higher redshifts and fainter sources.

\subsubsection{Rubin Observatory LSST}
The Rubin Observatory LSST will carry out an 18,000 deg$^2$ photometric galaxy survey at $0<z<4$ in optical bands \citep{2009arXiv0912.0201L}. The expected 5$\sigma$ point-source depth after a ten-year observation is a magnitude of 27.5 at $r$ band\footnote{\url{https://www.lsst.org/scientists/keynumbers}}
. A subset of galaxies from $0<z<3$ with a higher redshift accuracy ($\sigma_z/(1+z)<0.05$) constitute the ``gold sample,'' which has a depth of 25.3 in $i$ band \citep{2009arXiv0912.0201L}. We predict the number density of these two samples by applying these magnitude thresholds to our galaxy luminosity function model from \citet{2012ApJ...752..113H}.

Photometric redshift estimates are not available for $z>4$ sources from the Rubin Observatory LSST; nevertheless, redshift information of high-$z$ sources can be determined by the Lyman-break dropout technique. We thus estimate Lyman-break galaxy samples in a few broad redshift bands (listed in Table~\ref{T:surveys}) defined by the LSST filter boundaries and the redshifted Ly$\alpha$ wavelength (rest-frame $121.6$ nm). For these Lyman-break selected galaxies, we estimate their number density by applying the same magnitude threshold (27.5 in $r$ band) to our luminosity function model. We use Eq.~\ref{E:bI_conti} to estimate galaxy bias in both the photometric and Lyman-break samples.

\subsubsection{The Roman Space Telescope}
The planned photometric galaxy survey as part of the Roman Space Telescope HLS \citep{2018arXiv180403628D} covers 2200 deg$^2$ with a 5$\sigma$ depth of $\sim 26.5$ in $Y$, $J$, and $H$ bands \citep{2015arXiv150303757S}. We apply these magnitude thresholds to our IGL luminosity function model to predict the number density of the HLS photometric galaxies from $z=0$ to $z=6$. At $z>6$, we use the expected Lyman-break selected galaxy number density from the Roman Space Telescope Science Investigation Team EXPO (S. Finkelstein, private communication)\footnote{See slides from the presentation given by S. Finkelstein in the conference \textit{Astronomy in the 2020s: Synergies with WFIRST}, available at the time of writing at \url{https://www.stsci.edu/~dlaw/WFIRST2020s/slides/finkelstein.pdf}}. These number density values are consistent with the numerical simulation prediction from \citet{2016MNRAS.463.3520W} at $z\geqslant8$.
We also use Eq.~\ref{E:bI_conti} to estimate galaxy biases in both photometric and high-redshift samples. 

\section{Results}\label{S:results}
\subsection{An Example Cross Power Spectrum}

\begin{figure*}[ht!]
\begin{center}
\includegraphics[width=\linewidth]{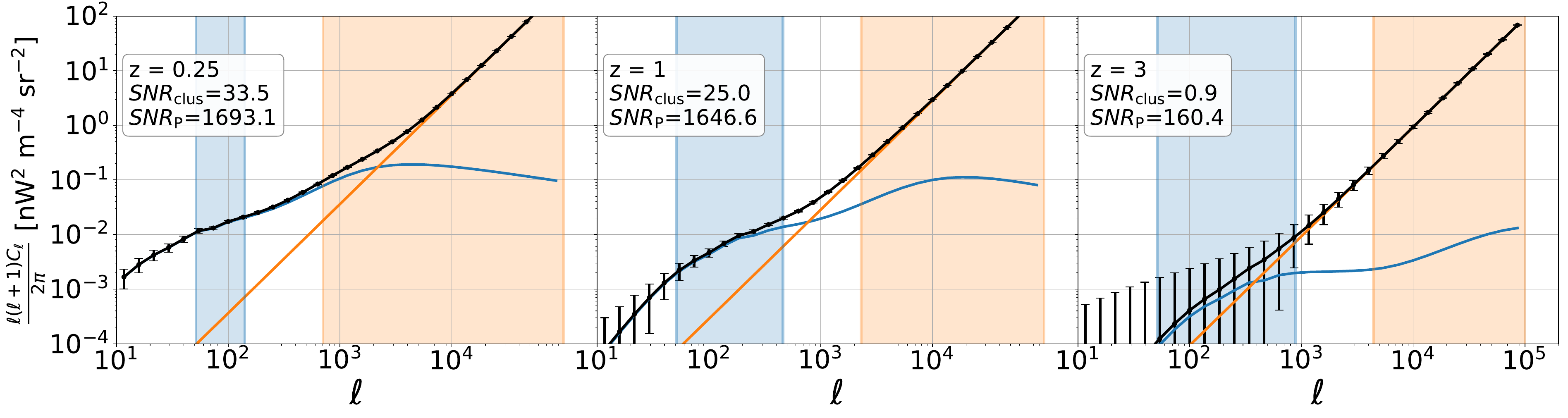}
\caption{\label{F:PS} Examples of cross power spectra (black) of the SPHEREx 3 $\mu$m image and SPHEREx spectroscopic galaxy catalog at $z=0.25$, $1$, and $3$. Blue and orange lines are the clustering and Poisson noise components, respectively. We constrain the redshift evolution of the combination of bias and redshift-dependent intensity of the EBL and of tracer galaxies, $b_I(z)d(\nu I_\nu)/dz(z,\nu)$ and $\left.d(\nu I_\nu)/dz\right|_g(z,\nu)$, respectively, from the cross power spectrum amplitudes in the clustering (blue band) and Poisson regimes (orange band), respectively; their SNR values are given in the legend. See Sec.~\ref{S:metrics} for details.}
\end{center}
\end{figure*}

In Fig.~\ref{F:PS}, we show a set of example cross power spectra of an all-sky SPHEREx intensity map at 3 $\mu$m and SPHEREx spectroscopic catalog at $z=0.25$, $1$, and $3$, respectively. The cross power spectra are given by  Eq.~\ref{E:Clx} with error bars estimated by  Eq.~\ref{E:delta_Clx}.  In the Poisson noise regime, we expect high-significance measurements of the cross power spectra at all three redshifts due to the large number of modes on small scales, while in the clustering regime, we expect significant detection at the lower two redshifts only. For the $z=3$ case, the low tracer number density gives a high galaxy Poisson noise level that reduces sensitivity on the cross power spectrum (see Sec.~\ref{S:error} for discussions).

We aim to extract two quantities related to the redshift-dependent IGL spectra, $b_I(z)d(\nu I_\nu)/dz(z,\nu)$ and $\left.d(\nu I_\nu)/dz\right|_g(z,\nu)$, from the large- (clustering) and small- (Poisson noise) scale amplitudes of the cross power spectrum. Their respective SNRs are denoted as ${\rm SNR}_{\rm clus}$ and ${\rm SNR}_{\rm P}$, respectively. Note that for ${\rm SNR}_{\rm P}$, we also include the non-Gaussian effect from the trispectrum term (Eq.~\ref{E:SNR_shot}). 

\subsection{Sensitivity to the Clustering Term}
\begin{figure*}[ht!]
\begin{center}
\includegraphics[width=\linewidth]{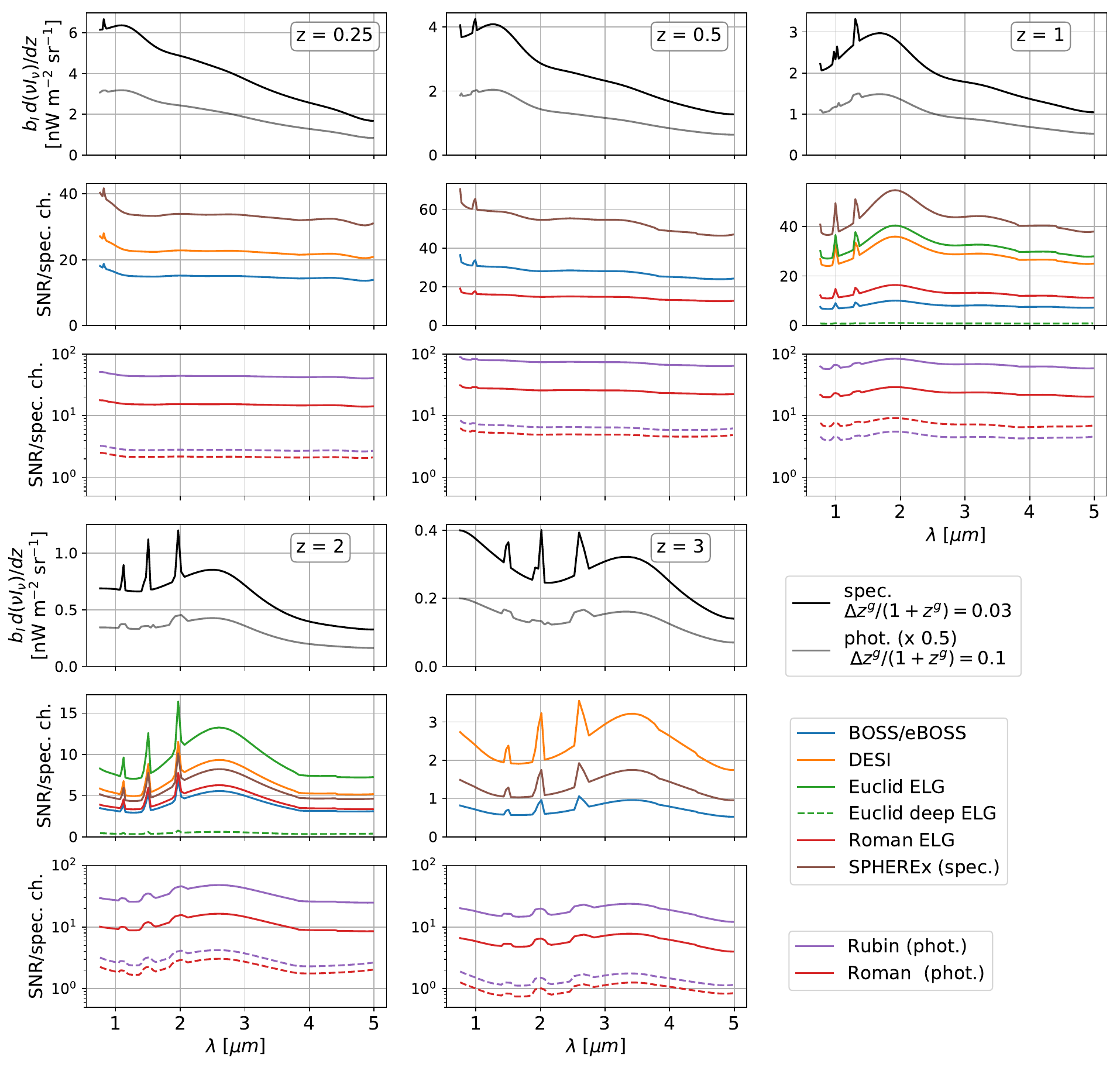}
\caption{\label{F:clus_spec} Constraints on the amplitude of cross power spectra of SPHEREx and galaxy surveys on large scales at $z\leqslant 3$. Top: $b_I(z)d(\nu I_\nu)/dz(z,\nu)$ from our model using spectroscopic (black) and photometric (gray, shifted down by $\times 0.5$ for presentation purposes) galaxy samples. The photometric surveys have wider redshift bins and thus lower spectral resolution. Middle: SNR forecast on $b_I(z)d(\nu I_\nu)/dz(z,\nu)$; (${\rm SNR}_{\rm clus}$ in Eq.~\ref{E:SNR_clus}) per SPHEREx spectral channel from cross-correlating SPHEREx images with different spectroscopic surveys. For the surveys that have multiple tracer catalogs, we plot the one that gives the highest SNR values (BOSS CMASS and DESI BGS for $z=0.25$; BOSS CMASS for $z=0.5$; eBOSS QSO and DESI ELG for $z=1$; eBOSS QSO and DESI QSO for $z=2$ and $3$). Solid lines are the cases using the SPHEREx all-sky survey, and dashed lines are the cases with two SPHEREx deep fields, which have a lower instrument noise but a smaller sky coverage. Bottom: the same as the middle rows but with photometric surveys.}
\end{center}
\end{figure*}

The top row of Fig.~\ref{F:clus_spec} shows our model for $b_I(z)d(\nu I_\nu)/dz(z,\nu)$ as a function of the 96 SPHEREx spectral bins and at the five fiducial $z\leqslant 3$ redshifts and masking thresholds. This quantity can be derived from the amplitude of the cross power spectra on large scales. We consider both spectroscopic and photometric galaxy surveys, where we use different redshift bin widths ($\Delta z^g/(1+z^g)=$ 0.03 and 0.1 for spectroscopic and photometric, respectively) to account for the respective redshift accuracies. For cross-correlation with photometric galaxies, the SPHEREx spectral line features are smoothed out due to the wide redshift bins.

The middle and bottom rows of Fig.~\ref{F:clus_spec} show the SNR forecasts for $b_I(z)d(\nu I_\nu)/dz(z,\nu)$, i.e., ${\rm SNR}_{\rm clus}$ in Eq.~\ref{E:SNR_clus}, per SPHEREx spectral channel by cross-correlating SPHEREx maps with different spectroscopic (middle) and photometric (bottom) galaxy catalogs. For surveys that have multiple tracer catalogs, we plot the ones that give the highest SNR values. With spectroscopic galaxies, we expect to achieve an ${\rm SNR}_{\rm clus}> 5$ out to $z\sim2$. However, at $z=3$, due to the low number density of tracers, ${\rm SNR}_{\rm clus}\sim 1$--$3$. Photometric galaxies, by contrast, have a much higher source density and can reach an ${\rm SNR}_{\rm clus}$ of $\sim$ $10(5)$ with Rubin Observatory LSST (Roman Space Telescope HLS) at $z=3$. Dashed lines indicate results from SPHEREx deep fields, which have a lower instrument noise but a smaller sky coverage. According to our model, the SNR in the deep fields are much lower than for the all-sky cases, due to their small $f_{\rm sky}$ that results in a large sample variance noise contribution. Note that with our chosen redshift binning there are $\sim 37$ measurements at $z<2$ and $\sim9$ measurements at $2<z<3$.

\begin{figure*}[ht!]
\begin{center}
\includegraphics[width=\linewidth]{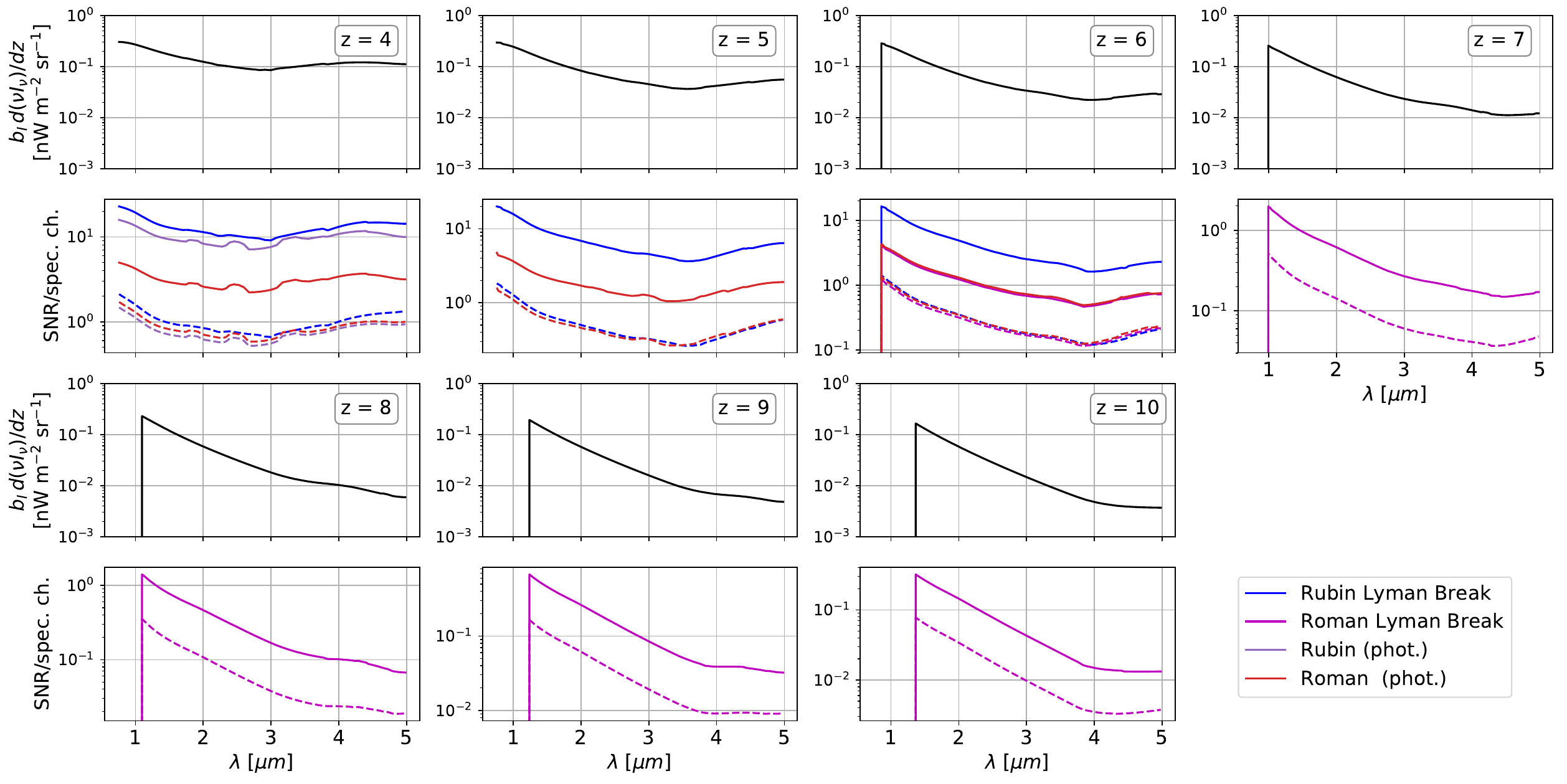}
\caption{\label{F:clus_phot} Constraints on the amplitude of cross power spectra of SPHEREx and galaxy surveys on large scales at $4<z<10$. Top: $b_I(z)d(\nu I_\nu)/dz(z,\nu)$ from our model using photometric surveys. We cut off the spectrum below the Lyman-$\alpha$ (121.6 nm) for $z>6$ since those high-energy photons will be absorbed by neutral gas in the intergalactic medium. Bottom: SNR forecast for $b_I(z)d(\nu I_\nu)/dz(z,\nu)$; (${\rm SNR}_{\rm clus}$ in Eq.~\ref{E:SNR_clus}) per SPHEREx spectral channel from cross-correlating SPHEREx images with different photometric surveys. The Lyman-break selected galaxy bins from Rubin Observatory LSST and the Roman Space Telescope are also shown in the panels close to their central redshifts. Solid lines are the cases using SPHEREx all-sky survey, and dashed lines are the cases with two SPHEREx deep fields.}
\end{center}
\end{figure*}

Fig.~\ref{F:clus_phot} shows the same forecast as Fig.~\ref{F:clus_spec} at high redshifts ($4<z<10$) with photometric surveys. Note with our chosen redshift binning there are $\sim 6$ measurements available between $3<z<6$. For $z>6$ where the universe has not yet fully ionized, we cut off the spectrum below the Lyman-$\alpha$ wavelength (121.6 nm rest frame) since those high-energy photons will be absorbed by the neutral gas in the intergalactic medium. 

The Lyman-break selected galaxy bins from Rubin Observatory LSST and Roman Space Telescope are also shown in the panels close to their central redshift. Note that the Lyman-break galaxy samples have a much wider redshift bin, and therefore they have a higher sensitivity but a lower spectral resolution on the inferred IGL spectrum.

Our estimate suggests that the IGL spectrum can be robustly measured over the full $0.75$--$5$ $\mu$m wavelength range across redshift, to a significance level of a few $\sigma$ per SPHEREx spectral channel out to $z=6$ by cross-correlating SPHEREx all-sky and upcoming photometric galaxy surveys. 

\subsection{Sensitivity to the Poisson noise Term}\label{S:results_shot}
\begin{figure*}[ht!]
\begin{center}
\includegraphics[width=\linewidth]{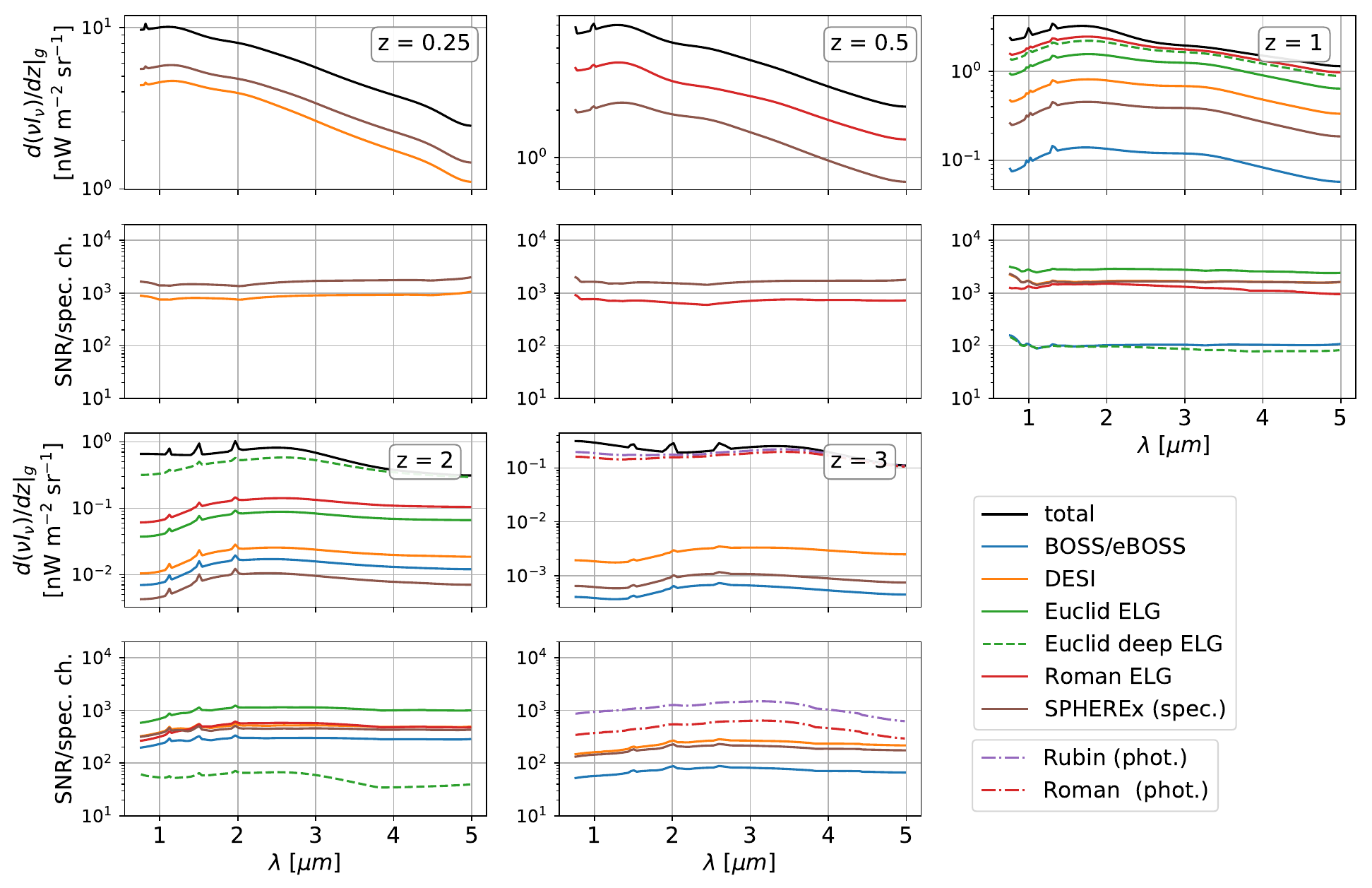}
\caption{\label{F:sh_spec} 
Constraints on the cross power spectrum Poisson noise amplitude at $z\leqslant 3$. Top: the IGL spectrum from all galaxies, $d(\nu I_\nu)/dz$ (black), and from the subset of the galaxies used for cross-correlations, $\left.d\nu I_\nu/dz\right|_g$ (colored), which is proportional to the Poisson noise amplitude. Solid lines give the cases using the SPHEREx all-sky survey, and dashed lines are the cases with the two SPHEREx deep fields, which have a lower instrument noise but a smaller sky coverage. Bottom:  SNR forecast for $\left.d(\nu I_\nu)/dz\right|_g$ (${\rm SNR}_{\rm P}$ in Eq.~\ref{E:SNR_shot}) per SPHEREx spectral channel from cross-correlating SPHEREx images with different spectroscopic surveys. For surveys that have multiple tracer catalogs, we use the same tracer as in Fig.~\ref{F:clus_spec}, which gives the highest SNR values. We use the fiducial masking depths as described in Sec.~\ref{S:fiducial_case}. In the CMASS case at $z=0.25$ and $z=0.5$, our chosen masking threshold is deeper than for the galaxy catalog ($m_{\rm th}>m_g$), and thus the cross Poisson power vanishes (see Eq.~\ref{E:dnuInu_dz_cont_g_0}). Therefore we only have clustering and no Poisson noise measurement for these two cases.} 
\end{center}
\end{figure*}

Fig.~\ref{F:sh_spec} shows our forecast on the cross power spectrum Poisson noise amplitude of SPHEREx and galaxy surveys at $z\leqslant 3$. The top row compares the redshift-dependent IGL spectrum from all sources, $d(\nu I_\nu)/dz$, and from the subset of sources in the galaxy samples used for cross-correlation, $\left.d(\nu I_\nu)/dz\right|_g$, which is proportional to the Poisson noise amplitude. Naturally, deeper catalogs capture a larger fraction of the total IGL in the cross Poisson noise amplitude. The bottom row shows our SNR forecast on $\left.d(\nu I_\nu)/dz\right|_g$, i.e. ${\rm SNR}_{\rm P}$, in Eq.~\ref{E:SNR_shot}. In most cases, the cross-correlation results in a highly significant ${\rm SNR}_{\rm P}$ ($\sim 10^2$--$10^4$) because of the large number of modes available on small scales. Dashed lines indicate the cross-correlation of the two SPHEREx deep field images with a Euclid deep field. Compared to all-sky, the deep fields capture a higher fraction of the IGL, as shown in the top row, but the SNR is lower due to the small sky coverage.

\begin{figure*}[ht!]
\begin{center}
\includegraphics[width=\linewidth]{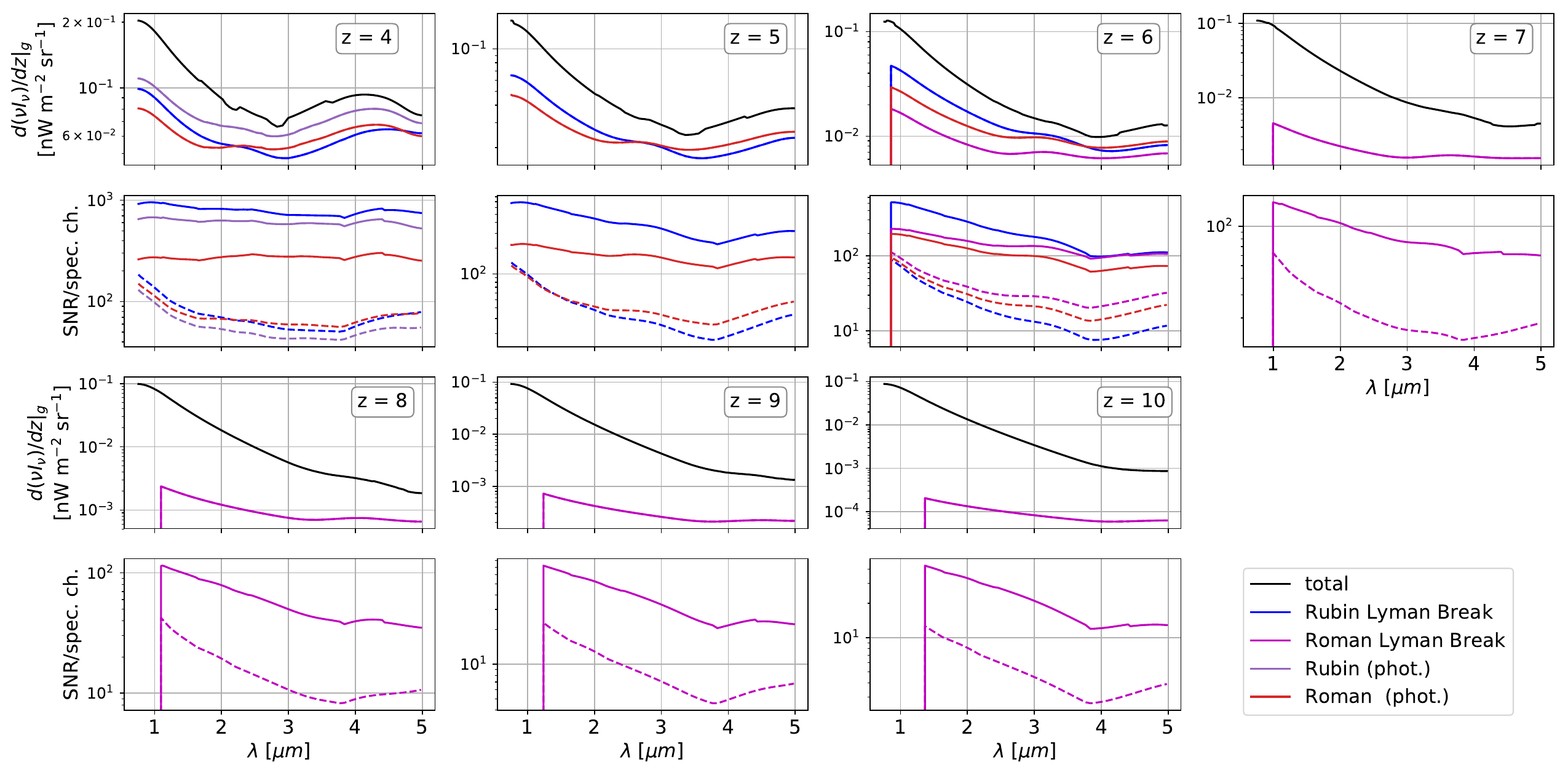}
\caption{\label{F:sh_phot} Constraints on the cross power spectrum Poisson noise amplitude for $4<z<10$. Top: the IGL spectrum from all galaxies, $d(\nu I_\nu)/dz$ (black), and from the subset of the galaxies used for cross-correlations, $\left.d\nu I_\nu/dz\right|_g$ (colored), which is proportional to the Poisson noise amplitude. Solid lines give the cases using the SPHEREx all-sky survey and dashed lines are the cases with the two SPHEREx deep fields, which have a lower instrument noise but a smaller sky coverage. The Lyman-break selected galaxy bins from Rubin Observatory LSST and Roman Space Telescope are also shown in the panels close to their central redshifts. We cut off the spectrum below the Lyman-$\alpha$ wavelength (121.6 nm) for $z>6$ since those high-energy photons will be absorbed by neutral gas in the intergalactic medium. Bottom: SNR forecast for $\left.d(\nu I_\nu)/dz\right|_g$ (${\rm SNR}_{\rm sh}$ in Eq.~\ref{E:SNR_shot}) per SPHEREx spectral channel from cross-correlations. The kink in SNR above $\lambda\sim 3.8$ $\mu$m is due to the change in SPHEREx resolving power and thus the instrument noise per channel (see Fig.~\ref{F:SPHEREx_noise}).}
\end{center}
\end{figure*}

Fig.~\ref{F:sh_phot} shows the same forecast as Fig.~\ref{F:sh_spec} at high redshifts ($4<z<10$) with photometric surveys and the Lyman-break selected samples. We observe a slight kink in SNR at $\lambda\sim 3.8$ $\mu$m due to the change in SPHEREx spectral resolution and thus the instrument noise level (see Fig.~\ref{F:SPHEREx_noise}). In practice, one can bin the measurements at $\lambda>3.8$ $\mu$m spectrally to increase the sensitivity. The same feature is not evident at the low-redshift cases as shown in Fig.~\ref{F:clus_spec}, because instrumental noise only becomes the dominant source of uncertainty in $\delta C_\ell^x$ (Eq.~\ref{E:delta_Clx}) at higher redshifts. Our forecast suggests that high-significance measurements with SNR $\sim 100$, can be achieved out to $z=6$ with upcoming photometric galaxy surveys.

Note that $\left.d(\nu I_\nu)/dz\right|_g$ is the mean spectrum of sources in the galaxy catalog, which is not the same as the mean IGL spectrum from the averaged emission from all sources. Nevertheless, for simplicity, we assume any subset of galaxies has the same mean spectrum as the IGL model used in this calculation. We also highlight potentially different measurements from the cross Poisson noise term with an example of quasars in Sec.~\ref{S:discussionQSO}.

\section{Discussion}\label{S:Discussion}
\subsection{Error Estimation}\label{S:error}
\begin{figure*}[ht!]
\begin{center}
\includegraphics[width=\linewidth]{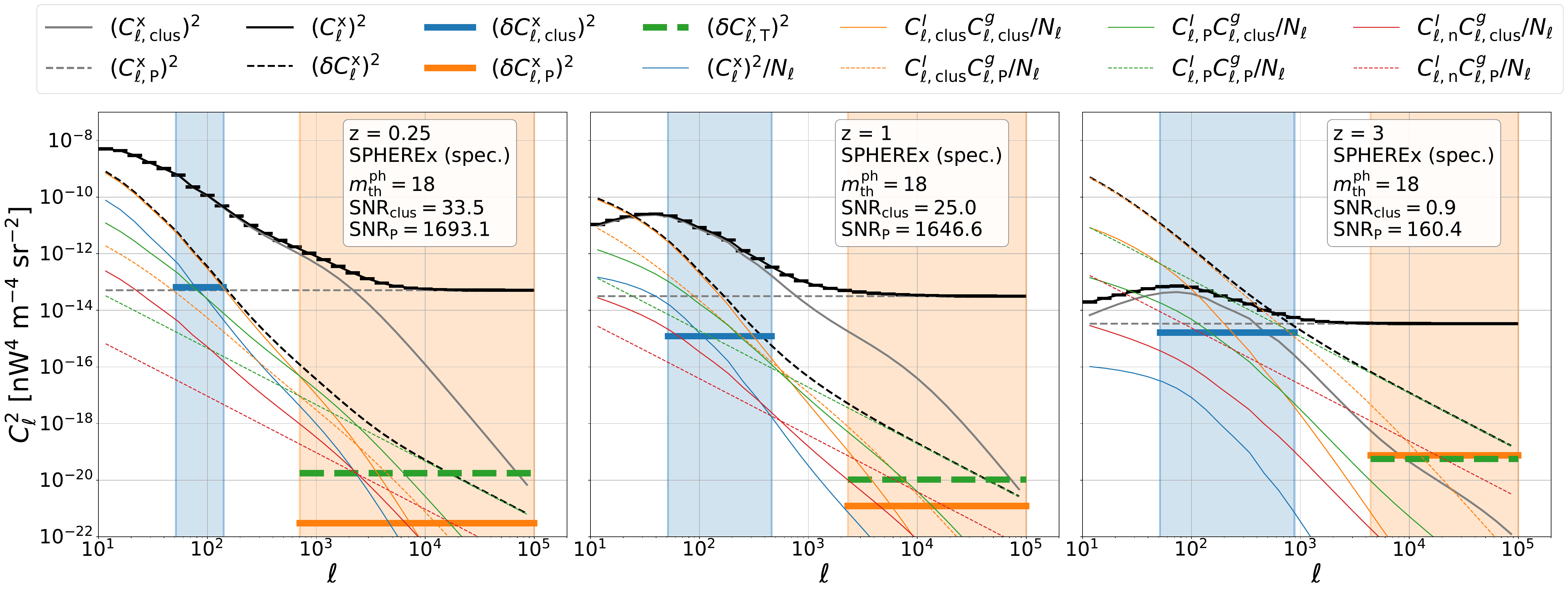}
\caption{\label{F:Cl_err_z} Cross power spectra (black solid lines) and their error components (colored lines) from cross-correlating the SPHEREx all-sky map at the 3$\mu$m channel with SPHEREx spectroscopic catalogs at $z=0.25$ (left), $1$ (middle), and $3$ (right), using a masking threshold of $m_{\rm th}^{\rm ph}=18$. 
Gray solid and dashed lines are the clustering and Poisson noise terms in the cross power spectrum, respectively. The thick black dashed lines show the total Gaussian noise power spectrum variance (i.e., the sum of all colored lines) as a function of $\ell$ bin, and the horizontal bars on the signal cross power spectra (short black solid lines) mark the $\ell$ binning. 
The thick blue and orange horizontal bars are the total Gaussian variance on the cross power spectrum clustering and Poisson noise amplitudes. They are computed by combining the multipole modes within the $\ell$ range of $[\ell^{\rm min}_{\rm clus},\ell^{\rm max}_{\rm clus}]$ (blue band) and $[\ell^{\rm min}_{\rm P},\ell^{\rm max}_{\rm P}]$ (orange band), respectively. The amplitude of the non-Gaussian variance from trispectrum is shown by the green dashed horizontal bar (see Eq.~\ref{E:SNR_clus} and \ref{E:SNR_shot}). The two lower-redshift cases are background-limited where the cross spectrum errors are dominated by $C_{\ell,\rm clus}^I C_{\ell,\rm clus}^g/N_\ell$, whereas at $z=3$, the sensitivity is limited by the galaxy Poisson noise due to its low tracer number density.}
\end{center}
\end{figure*}

The variance of a cross power spectrum is the sum of several contributing factors in Eq.~\ref{E:delta_Clx}, including instrument noise, galaxy Poisson (shot) noise, and sample variance.  Fig.~\ref{F:Cl_err_z} compares three cross power spectra and their different error components from cross-correlating SPHEREx all-sky maps at the 3$\mu$m channel with SPHEREx spectroscopic catalogs at $z=0.25$,  $1$, and $3$, using a masking threshold of $m_{\rm th}^{\rm ph}=18$. The Gaussian error on the cross power spectrum (Eq.~\ref{E:delta_Clx}) comprises a quadratic sum of individual power spectrum terms:
\begin{equation}
\begin{split}
\left ( \delta C_\ell^{\rm x} \right )^2 = \frac{1}{N_\ell}&\left [(C_{\ell}^{\rm x})^2+ C_{\ell}^I C_{\ell}^g \right ]\\
=\frac{1}{N_\ell}&\left [\right. (C_{\ell}^{\rm x})^2 +\left (C_{\ell, \rm clus}^I+C_{\ell, \rm P}^I + C_{\ell,n}^I \right )\\
&\cdot\left ( C_{\ell, \rm clus}^g+C_{\ell, \rm P}^g \right )  \left.\right ]\\
=\frac{1}{N_\ell}&\left [\right.(C_{\ell}^{\rm x})^2+C_{\ell, \rm clus}^IC_{\ell, \rm clus}^g+C_{\ell, \rm clus}^IC_{\ell, \rm P}^g\\
&+C_{\ell, \rm P}^IC_{\ell, \rm clus}^g+C_{\ell, \rm P}^IC_{\ell, \rm P}^g \\
&+ C_{\ell,n}^IC_{\ell, \rm clus}^g+ C_{\ell,n}^IC_{\ell, \rm P}^g\left.\right ].
\end{split}
\end{equation}

First, we see that the cross spectrum amplitude decreases with redshift following the redshift dependence of the IGL spectrum (Fig.~\ref{F:nuInu}) and the underlying matter density fluctuations. For the two lower-redshift cases, the dominant noise term is $C_{\ell,\rm clus}^I C_{\ell,\rm clus}^g/N_\ell$. While $C_{\ell,\rm clus}^g$ is determined by the underlying cosmology, the only way to further bring down the error on the cross power spectrum is by reducing the amplitude of $C_{\ell,\rm clus}^I$, which is dominated by aggregate emission along the line of sight. Therefore, these measurements are in what we refer to as the background-limited regime, where background here  refers to all emission not from the redshift of interest (and thus includes the foreground, too). We can apply a deeper mask to suppress the background (foreground). However, this will also reduce the signal, and therefore there exists a masking threshold that optimizes the trade-off between background (foreground) and signal removal. We detail this comparison in Sec.~\ref{S:masking}.

At $z=3$, the dominant noise term becomes $C_{\ell,\rm clus}^I C_{\ell,\rm P}^g/N_\ell$ due to the increased galaxy Poisson noise power spectrum amplitude, $C_{\ell,\rm P}^g$, in available surveys. This means that a deeper catalog with a higher tracer number density can effectively improve the sensitivity.

\begin{figure*}[ht!]
\begin{center}
\includegraphics[width=\linewidth]{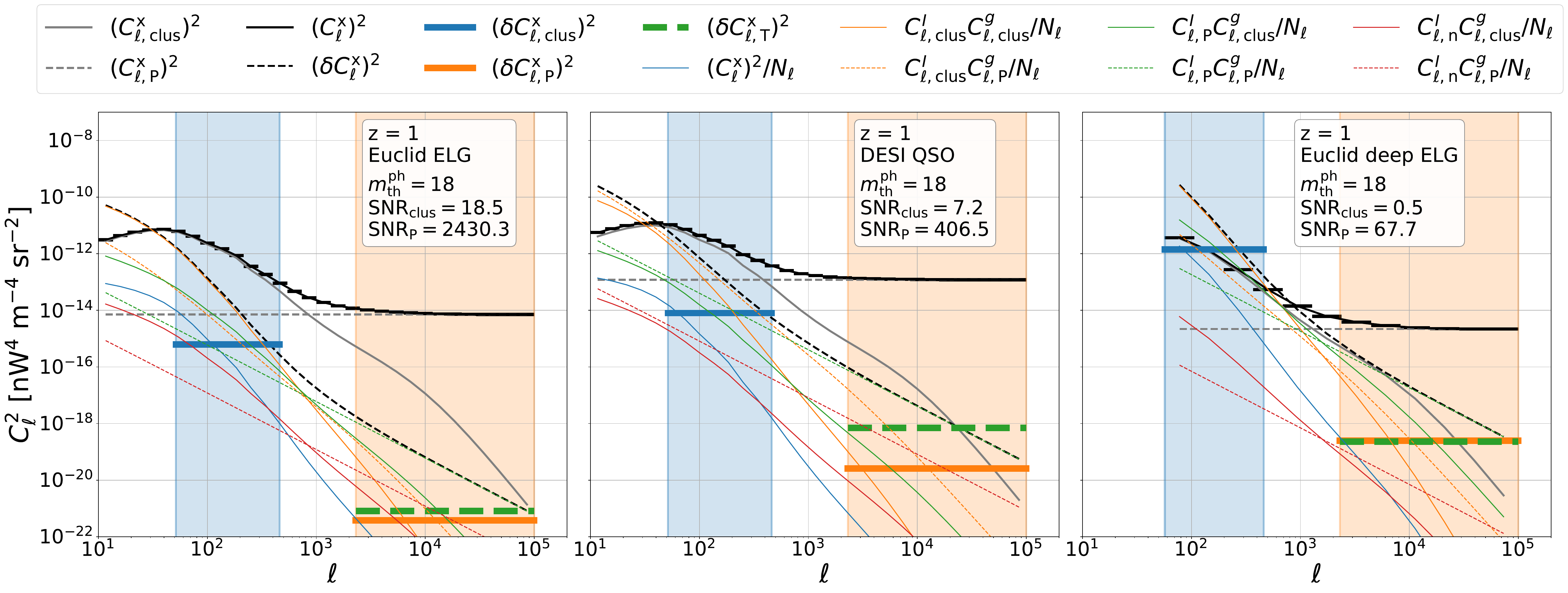}
\caption{\label{F:Cl_err_ng} Cross power spectra (black solid lines) and their error components (colored lines) from cross-correlating the SPHEREx all-sky map at the 3$\mu$m channel with Euclid ELG (left) and DESI QSO (middle), and the SPHEREx deep field map at 3$\mu$m with the Euclid deep field ELG catalog (right). All three cases are at $z=1$ with a masking threshold of $m_{\rm th}^{\rm ph}=18$. The curves are the same as in Fig.~\ref{F:Cl_err_z}. The Euclid ELG (left) and DESI QSO (middle) are comparable in survey area size, but the former has a lower tracer number density and thus a higher Poisson noise that suppresses the sensitivity. The Euclid deep field case (right) has a higher ELG number density than the wide-field case (left), however it has a lower SNR due to the larger sample variance from its smaller sky coverage.}
\end{center}
\end{figure*}

Fig.~\ref{F:Cl_err_ng} compares three different galaxy surveys at $z=1$ with the same masking threshold of $m_{\rm th}^{\rm ph}=18$ for the SPHEREx 3$\mu$m channel. The first two, the Euclid ELG and DESI QSO surveys, have a similar survey area, $f_{\rm sky}$, and thus similar $N_\ell$'s, but Euclid ELG has a higher source number density. As a result, the two cases have similar cross spectrum amplitudes in the clustering regime, but DESI QSO has a higher Poisson noise amplitude, and the noise terms that contain $C_{\ell, \rm P}^g$ (all dashed lines) are therefore higher in the DESI QSO case. 

We can also compare the Euclid wide and deep field cross-correlations with the SPHEREx all-sky and deep fields, respectively. As shown on the left and right panels of Fig.~\ref{F:Cl_err_ng}, we see that the terms associated with galaxy Poisson noise (all dashed lines) are lower in the deep fields because of the higher galaxy number density. However, the deep fields cannot access the lower $\ell$ modes, and the overall noise is higher in the deep field due to the larger sample variance (small $N_\ell$) from its smaller sky coverage.

\begin{figure*}[ht!]
\begin{center}
\includegraphics[width=\linewidth]{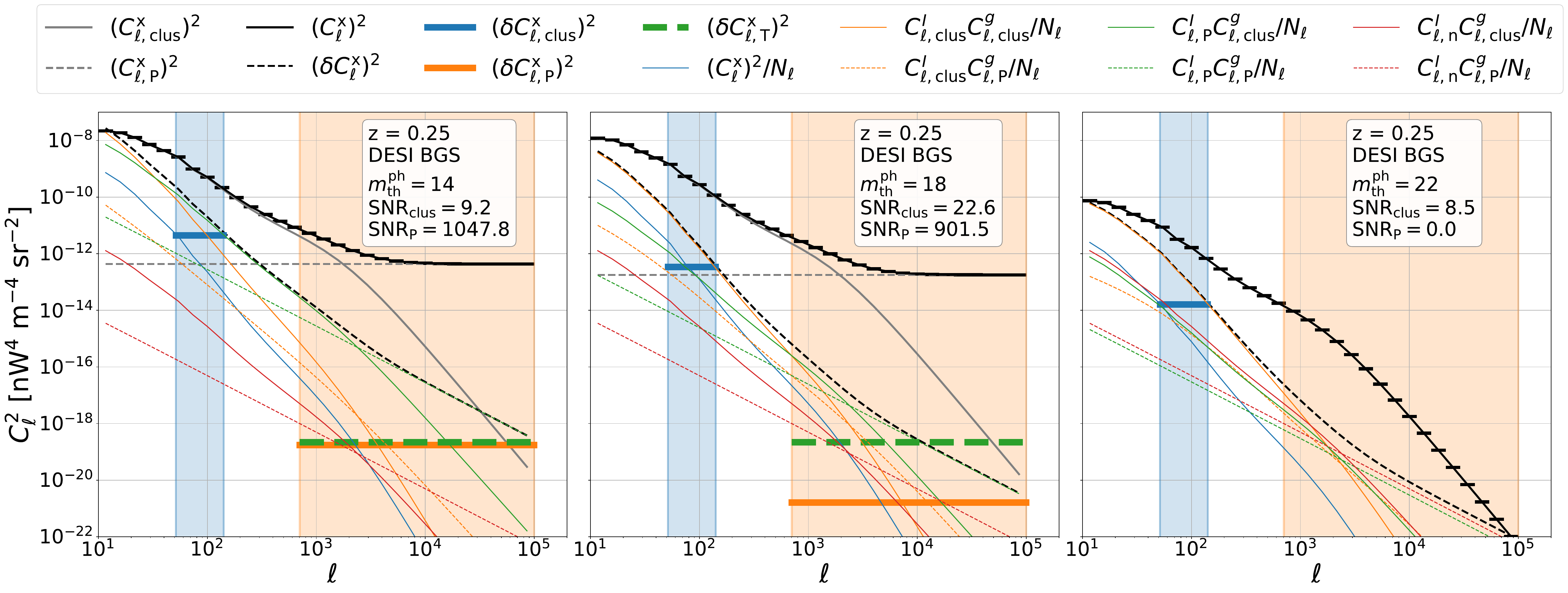}
\caption{\label{F:Cl_err_mask} Cross power spectra (black solid lines) and their error components (colored lines) from cross-correlating the SPHEREx all-sky map at the  3$\mu$m channel with 
DESI BGS at $z=0.25$ using a masking threshold of $m_{\rm th}^{\rm ph}=14$ (left), $18$ (middle), and $22$ (right), respectively. The curves are the same as in Fig.~\ref{F:Cl_err_z}. The SNR is improved by raising the masking threshold from $m_{\rm th}^{\rm ph}=14$ to $m_{\rm th}^{\rm ph}=18$, since the line-of-sight contamination (the $C_{\ell, \rm clus}^{\rm I}$ and $C_{\ell, \rm P}^{\rm I}$ terms) decreases, while signal reduction is negligible (black solid lines). However, if we further increase the masking threshold to $m_{\rm th}^{\rm ph}=22$, the SNR becomes lower due to the significant loss of signal by masking.}
\end{center}
\end{figure*}

Fig.~\ref{F:Cl_err_mask} demonstrates the impact of masking depth, using the cross-correlation of  SPHEREx all-sky maps at 3 $\mu$m and DESI BGS galaxies at $z=0.25$ with three different masking depths, $m_{\rm th}^{\rm ph}=14, 18, 22$. Comparing $m_{\rm th}^{\rm ph}=14$ and $m_{\rm th}^{\rm ph}=18$, the signals are almost the same (black solid line) but the total Gaussian error (black dashed line) is lower with a deeper mask; this suggests that by masking sources to $14<m_{\rm th}^{\rm ph}<18$, a significant portion of the foreground can be removed, while the signal reduction is negligible, which leads to an improvement of SNR. However, if we further increase the masking depth to $m_{\rm th}^{\rm ph}=22$, the signal is drastically reduced, which means a significant fraction of IGL at this redshift is from galaxies with $18<m_{\rm th}^{\rm ph}<22$, and we cannot get a higher SNR with this deeper masking depth.

\subsection{Masking Depth}\label{S:masking}
\begin{figure*}[ht!]
\begin{center}
\includegraphics[width=\linewidth]{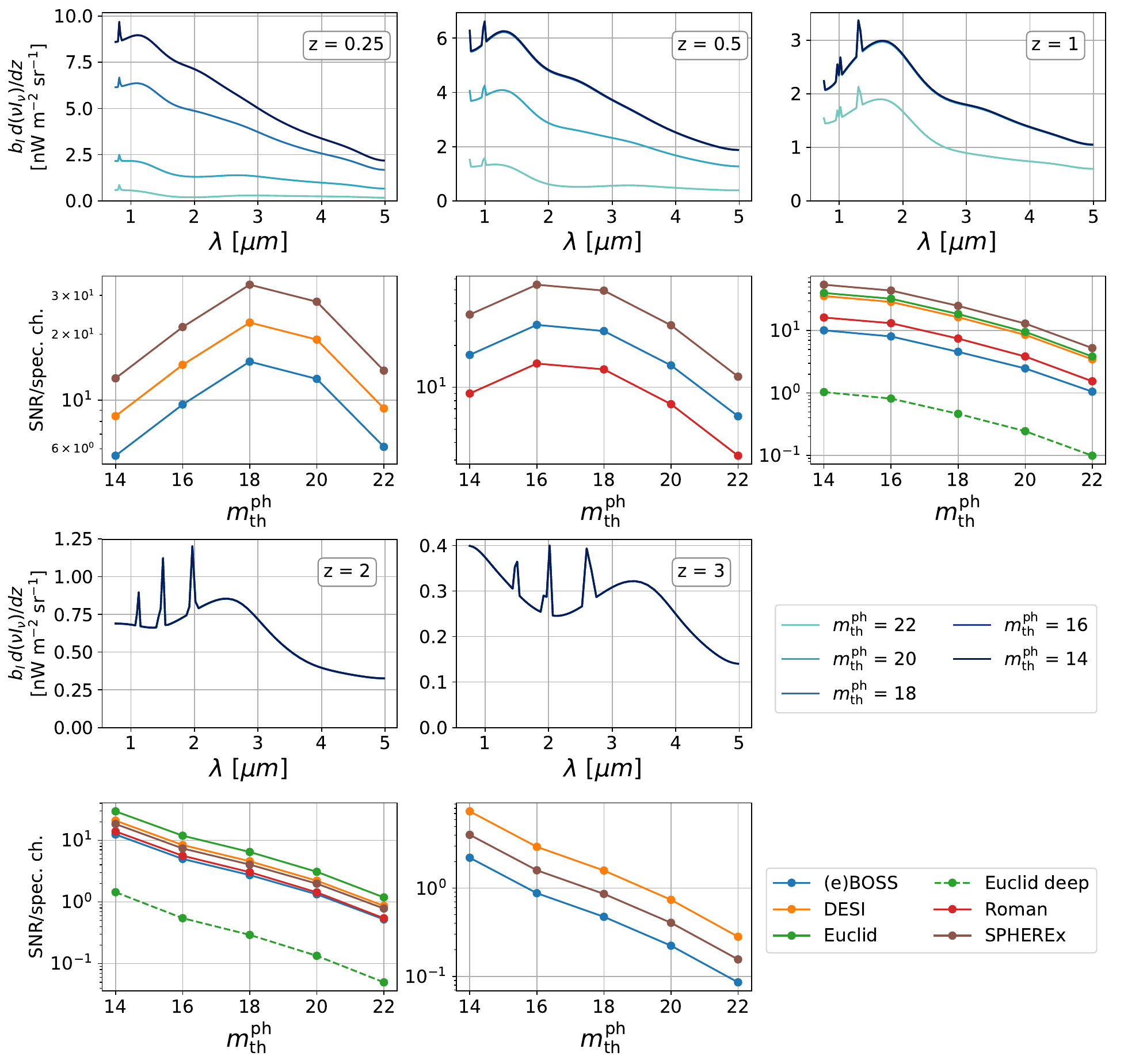}
\caption{\label{F:clus_mth} Top: the IGL spectrum $d(\nu I_\nu)/dz$ in the SPHEREx 3$\mu$m channel as a function of masking depth $m_{\rm th}^{\rm ph}$ at fiducial masking wavelength $\lambda^{\rm ph}=1$ $\mu$m. Bottom: ${\rm SNR}_{\rm clus}$ as a function of $m_{\rm th}^{\rm ph}$ with different galaxy surveys shown in Fig.~\ref{F:clus_spec}.}
\end{center}
\end{figure*}

We consider masking all sources from an external catalog that are brighter than a flux threshold, and therefore the choice of masking depth is a trade-off between signal loss and foreground reduction. A deeper masking depth tends to remove more foreground sources contributing to the intensity maps, but at the same time, bright galaxies in the target redshift range of interest can also be masked. Fig.~\ref{F:clus_mth} demonstrates how the IGL amplitude and the error depends on masking depth. Here, we consider 
masking sources brighter than magnitude threshold $m_{\rm th}^{\rm ph}$ at the fiducial masking wavelength $\lambda^{\rm ph}=1$ $\mu$m. At lower redshift, the IGL amplitude strongly depends on the masking depths $m_{\rm th}^{\rm ph}$. At higher redshift, the signals are almost unchanged with varying masking depth, as the sources that dominate the IGL at these redshifts are below the masking thresholds. In other words, at low redshift, there is a masking depth that optimizes the trade-off between signal loss and foreground reduction. At higher redshift, deeper masks perform better since they reduce more foreground emission while having a negligible impact on the signal. 

Here, we only consider masking all stars and galaxies brighter than a magnitude threshold. In practice, if the external catalog provides redshift information of individual sources, a redshift-dependent magnitude threshold might also help to minimize foreground contamination from lower-redshift clustering signals. We leave this investigation to future work.

We ignore the effect of pixel loss due to masking in our forecast. In reality, masking results in highly nonuniform images, which not only reduces the number of multipole modes in the data, but also introduces mode mixing, the coupling of signal on different scales that needs to be corrected for in analysis. The reduction in power spectral sensitivity due to masking as a function of multipole depends on the source luminosity function and clustering amplitude, the masking size around each source, and the mode-coupling correction method. A more detailed study requires realistic simulated images, which is beyond the scope of this work.

\subsection{Caveats of the Abundance Matching Model}\label{S:discussion_caveats}
In our model, we use the abundance matching technique to relate the underlying halo mass distribution to the luminosity of galaxy tracer continuum and line emissions. In other words, if we have $N$ galaxy tracers, we assume they are in the most massive $N$ dark matter halos, and are also the strongest emitters of continuum and spectral lines in all frequencies. In reality, the relation has a large scatter. It has been shown that red and blue galaxies have different spatial clustering amplitudes and trace the matter density field differently \citep{2011ApJ...736...59Z}. Each galaxy has a distinct color and spectral line strength depending on its stellar composition, metallicity, dust attenuation, and other interstellar medium properties. Moreover, the tracer galaxies are typically selected by certain spectral features, making them unrepresentative of the average EBL emitters. For example, the ELG samples will bias toward star-forming galaxies, which have strong emission lines. The decorrelation between matter and continuum and line emissions will not only introduce uncertainties to our model, but also reduce the cross-correlation signal on small scales. As discussed in \citet{2019ApJ...877..150C}, the stochasticity in the relation among mass, tracers, and EBL emission could be a possible explanation of their low EBL intensity bias values. Quantifying these decorrelations requires a coherent model for mass and light, which we leave to future work.

\subsection{Intensity Bias Degeneracy}\label{S:bias}
We derive constraints on $b_I(z)d(\nu I_\nu)/dz(z,\nu)$ from the cross power spectrum amplitude in the clustering regime, but we do not further consider decoupling these two terms. In practice, we can apply the methods used in \cite{2019ApJ...877..150C} to jointly fit the redshift and frequency dependencies of bias and intensity, using all the cross spectrum measurements. In \cite{2019ApJ...877..150C}, they calibrate the total intensity to the intensity of resolved sources observed in the local universe, as the sources are sufficiently faint to account for most of the intensity from all emitters. In our case, we can cross-correlate SPHEREx maps with a deep catalog at low redshift, and similarly assume $\left.d(\nu I_\nu)/dz\right|_g\approx d(\nu I_\nu)/dz$ to break the bias-intensity degeneracy. According to our model, SPHEREx spectroscopic catalog at $z=0.1$ contains $\sim 95\%$ of the total IGL intensity, and thus this approach is feasible for our interpretation, assuming little redshift evolution of the EBL composition.



\subsection{Other Foreground Components}
In this work, we only consider Galactic stars as the foreground source and estimate their noise contribution to the cross power spectrum. In practice, the SPHEREx intensity measurements are also contaminated by the zodiacal light, diffuse Galactic light, Galactic dust and polycyclic aromatic hydrocarbon (PAH) emissions. These local foreground emissions will not spatially correlate with tracers of large-scale structure considered in this work; however, they may contribute to the noise level of the cross power spectrum if untreated in the data analysis. Zodiacal light is the brightest emission in the near-infrared sky, caused by sunlight scattered by interplanetary dust. It is smooth on degree scales and can be mitigated by a high-pass filter in map space, and is therefore not expected to contribute to the cross power spectrum noise. The diffuse Galactic light, starlight scattered by interstellar dust, has fluctuations on degree scales with a known angular power spectrum. It is highly correlated with the far-infrared thermal emission from dust grains in the optically thin limit \citep{2013ApJ...767...80I}, and therefore can be estimated and largely mitigated by scaling the far-infrared images to reduce its impact on the cross-power spectrum noise. Finally, at SPHEREx's longer wavelength bands ($\gtrsim 3$ $\mu$m), the Galactic dust and PAH emission from the interstellar medium also contribute to additional sky brightness. We have excluded the Galactic plane regions and limit our analysis to 75\% of the sky throughout, and expect their contributions to cross-power spectrum noise to be small. A detailed analysis is warranted but is beyond the scope of this study, and we leave further investigations to future work.

\section{Science Interpretation}\label{S:Interpretation}

\subsection{Poisson Noise Constraints on the Tracer Spectrum}\label{S:discussionQSO}
\begin{figure}[ht!]
\begin{center}
\includegraphics[width=\linewidth]{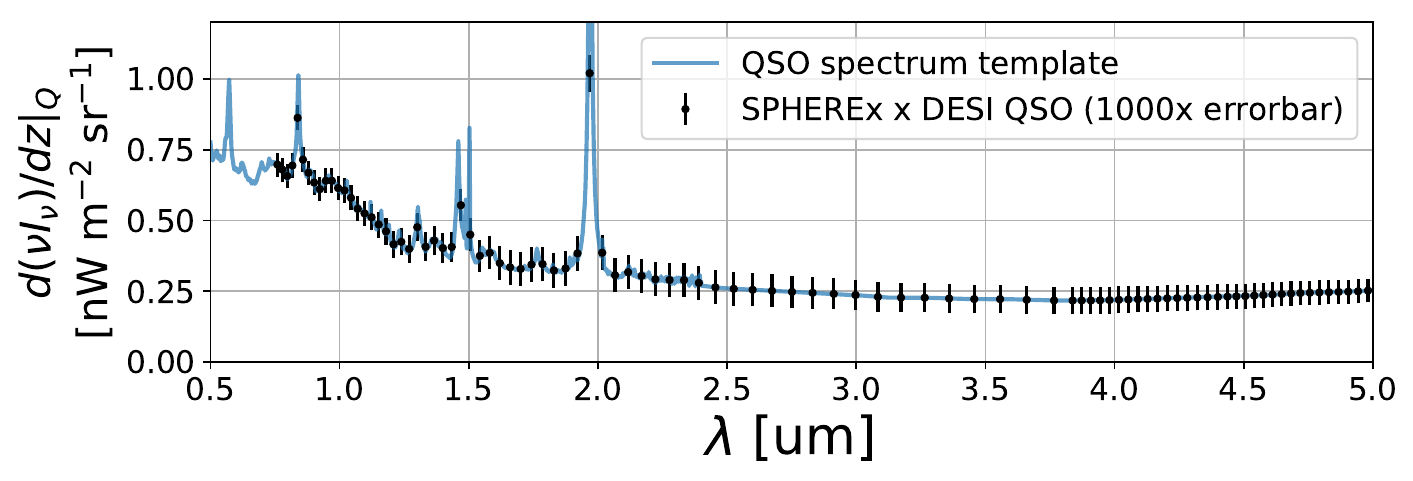}
\caption{\label{F:QSO} Average quasar spectrum constraints (per SPHEREx spectral bin) from cross-correlating SPHEREx with DESI quasar samples at $z=2$. The blue line is the modeled quasar spectrum template from \citet{2018AJ....156...66M}.}
\end{center}
\end{figure}

In the Poisson noise sensitivity forecast presented in Sec.~\ref{S:results_shot}, we assume the tracer population has the same averaged spectrum as our IGL model spectrum. This is not necessarily true if we select a certain type of source for cross-correlation (e.g., ELG, QSO), where the Poisson noise signal will be the averaged spectrum of the tracer sources. Therefore we can extract potentially distinct information from the Poisson noise spectrum. As a demonstration, we consider a case of cross-correlating SPHEREx with quasars from DESI at $z=2$, using a quasar spectrum and luminosity function model described in Sec.~\ref{S:QSO}. The constraints on the averaged quasar spectrum, $\left. d(\nu I_\nu)/dz\right |_Q$, from the cross Poisson spectrum is shown in Fig.~\ref{F:QSO}. Using the small-scale cross-correlation information of SPHEREx with the quasar sample, we can obtain a high-SNR measurement on the average quasar spectrum at near-infrared wavelengths across redshift.

Given the high-SNR value from the Poisson noise measurements in all cases considered in this work, it is also feasible to gain more information from Poisson noise by splitting the tracers into sub-samples by different properties. For example, we can select galaxies by their stellar mass, star formation rate, and morphology, and constrain their average spectrum individually.

\subsection{Science from Nonlinear Clustering}
In this work, we only consider information from the linear and Poisson noise scales of the cross power spectrum. Nonlinear clustering scales also contain crucial information on galaxy evolution as well as the underlying cosmology. For example, emission from satellite galaxies and the contribution of IHL can be constrained by the amplitude and shape of the nonlinear power spectrum \citep[e.g., ][]{2012Natur.490..514C, 2014Sci...346..732Z,2021arXiv210303882C}. As we have demonstrated that a high-SNR measurement can be achieved by cross-correlating SPHEREx with current or upcoming galaxy surveys, it is promising to extract more constraints from nonlinear scales.

\subsection{Science from Near-infrared EBL Tomography}
The rest-frame near-infrared light from galaxies are dominated by low-mass stars, and thus the near-infrared photometry has been used to measure the stellar mass function across redshifts \citep{2014ARA&A..52..415M}. Most of the current constraints on the stellar mass function are based on deep photometric samples \citep{2013ApJ...777...18M,2018MNRAS.480.3491W,2021MNRAS.505..540T}. While individual galaxies can be resolved down to the $\sim 10^{-2}$--$10^{-3}$ $M*$ mass limit at lower redshift \citep[e.g. ][]{2013ApJ...777...18M}, at $z\gtrsim 3$, photometric surveys are only complete down to $\sim M*$ scales, and the faint end estimation are extrapolated from the bright end with a functional fit \citep{2021MNRAS.505..540T}. SPHEREx EBL measurement, $d\nu I_\nu/dz(\nu,z)$, provides complementary information to the current galaxy detection-based measurement. Although individual galaxy spectra cannot be extracted from the diffuse maps, SPHEREx EBL measurements contain emission from all sources with high spectral and redshift resolutions, and cover a large sky area. They can be used to calibrate the stellar mass function model built from the resolved galaxies, and probe emission from diffuse origins such as the IHL. At higher redshift, SPHEREx covers the rest-frame optical and ultraviolet spectra and helps constrain the star formation history \citep{2014ARA&A..52..415M,2014arXiv1412.4872D, 2018arXiv180505489D}.

\subsection{Science from Spectral Line Intensity Mapping}
While quoting the SNR, we do not separate the significance of spectral lines from continuum emission in our results. Our forecast suggests that spectral lines can be detected with a high significance at lower redshift ($z\lesssim 3$) in both the clustering and Poisson noise regimes. Line emissions contain valuable information on the galaxy properties. For example, the H$\alpha$, Ly$\alpha$, and [\ion{O}{ii}] line flux can be used to infer the star formation rate \citep{2004MNRAS.351.1151B,2013MNRAS.433.2764G,2013MNRAS.428.1128S,2015MNRAS.452.3948K}; the Balmer decrement, the ratio of H$\alpha$ and H$\beta$ line flux, is a crucial indicator of dust extinction \citep{2013ApJ...763..145D}. The SPHEREx cross-correlation measurements of line emissions in the intensity mapping regime have the potential to address these galaxy properties across redshifts. Moreover, with the high spectral resolution and sensitivity, SPHEREx offers a unique probe of the three-dimensional large-scale structure with line intensity mapping \citep[e.g. ][]{2017arXiv170909066K} for cosmological constraints in the high-redshift universe.

\subsection{Cosmological Constraints}
So far we have focused on EBL spectrum constraints derived from cross-correlation, which only 
depend on the large-scale and Poisson noise amplitude of the cross power spectrum. In general, abundant cosmological information can be inferred from the power spectrum; for example, baryon acoustic oscillation signals can be extracted from the cross power spectrum, which can then be used to constrain the growth and geometry of the universe \citep{2005ApJ...633..560E}. 

\section{Conclusion}\label{S:Conclusion}
We forecast the sensitivity of near-infrared EBL tomography using spectro-imaging from the upcoming SPHEREx mission through cross-correlation with several current and future galaxy surveys, spanning redshifts of $ 0.25 < z < 10 $. We consider IGL as the only emitting source that constitutes the near-infrared EBL, and build a model for IGL continuum and spectral line emissions. We then model the cross power spectrum of SPHEREx images and galaxy surveys as a function of SPHEREx spectral channel and galaxy redshift. 

From the amplitudes of the cross power spectrum on linear scales, we infer the redshift-dependent EBL spectrum, multiplied by a bias factor, $b_I(z)d(\nu I_\nu)/dz(z,\nu)$. Our forecast suggests that this quantity can be constrained to a significance of at least a few $\sigma$ out to $z=6$, using the SPHEREx all-sky survey in cross-correlation with a combination of spectroscopic (at $z\lesssim 3$) and photometric galaxy surveys (at $ 0.25 < z < 10$). At $3\lesssim z\lesssim 10$, photometric galaxy catalogs from the upcoming Rubin Observatory and the Roman Space Telescope can constrain $b_I(z)d(\nu I_\nu)/dz(z,\nu)$, albeit at a lower redshift and spectral resolution. 

The Poisson noise level of the cross power spectrum represents the average intensity of sources used in cross-correlation. Our forecast shows that the Poisson noise level can be extracted at a high significance ($\gtrsim 10^2$) out to $z\sim10$, suggesting that the SPHEREx all-sky and deep field data can provide high-sensitivity measurements of the average near-infrared spectrum of any selected population of tracer sources.

In summary, a high-sensitivity tomographic measurement of the EBL spectrum can be achieved by cross-correlating SPHEREx with current and future galaxy spectroscopic and photometric surveys. Making use of the clustering and Poisson noise of the cross power spectra, we expect further cosmological and astrophysical information can be extracted from this rich dataset.

\acknowledgments
We are grateful to Jamie Bock, Yi-Kuan Chiang, Emmanuel Schann, Olivier Dor{\'e}, Richard Feder, Lluis Mas-Ribas, Daniel Masters, and Rogier Windhorst for helpful discussions and comments on the draft. We are in debt to Lluis Mas-Ribas for providing the quasar spectrum template, Emmanuel Schann for suggesting the trispectrum correction, and Steven Finkelstein for providing the Roman Space Telescope High Latitude Survey galaxy number count estimates. Y.-T.C. acknowledges
support by the Ministry of Education, Taiwan through the Taiwan-Caltech Scholarship. T.-C.C. acknowledges support from the JPL Strategic R\&TD awards. Part of this work was done at the Jet Propulsion Laboratory, California Institute of Technology, under a contract with the National Aeronautics and Space Administration.

\bibliography{reference}{}
\bibliographystyle{aasjournal}

\end{document}